\documentclass[prb, amsmath, amssymb]{revtex4}
\usepackage{graphicx}
\usepackage{float}
\pdfoutput=1
\begin{document}
\title{Gaussian core model phase diagram and pair correlations in high Euclidean dimensions}
\author{Chase E. Zachary}
\email{czachary@princeton.edu}
\author{Frank H. Stillinger}
\email{fhs@princeton.edu}
\affiliation{Department of Chemistry, Princeton University, Princeton, New Jersey 08544, USA}
\author{Salvatore Torquato}
\email{torquato@electron.princeton.edu}
\affiliation{Department of Chemistry, Princeton University, Princeton, New Jersey 08544, USA;\\
Program in Applied and Computational Mathematics, Princeton University, Princeton, New Jersey 08544, USA;\\
Princeton Institute for the Science and Technology of Materials, Princeton University, Princeton, New Jersey 08544, USA;\\
Princeton Center for Theoretical Physics, Princeton University, Princeton, New Jersey 08544, USA}

\date{\today}

\begin{abstract}
The physical properties of a classical many-particle system with interactions given by a repulsive Gaussian pair potential are extended to arbitrarily high Euclidean dimensions.  The goals of this paper are to characterize the behavior of the pair correlation function $g_2$ in various density regimes and to understand the phase properties of the Gaussian core model (GCM) as parametrized by dimension $d$.  To this end, we explore the fluid (dilute and dense) and crystalline solid phases.  For the dilute regime of the fluid phase, a cluster expansion of $g_2$ in reciprocal temperature $\beta$ is presented, the coefficients of which may be evaluated analytically due to the nature of the Gaussian potential.  We present preliminary results concerning the convergence properties of this expansion.  The analytical cluster expansion is related to numerical approximations for $g_2$ in the dense fluid regime by utilizing hypernetted chain, Percus-Yevick, and mean-field closures to the Ornstein-Zernike equation.  Based on the results of these comparisons, we provide evidence in support of a decorrelation principle for the GCM in high Euclidean dimensions.  In the solid phase, we consider the behavior of the freezing temperature $T_f (\rho)$ in the limit $\rho \rightarrow +\infty$ and show $T_f (\rho) \rightarrow 0$ in this limit for any $d$ via a collective coordinate argument.  Duality relations with respect to the energies of a lattice and its dual are then discussed, and these relations aid in the Maxwell double-tangent construction of phase coexistence regions between dual lattices based on lattice summation energies.  The results from this analysis are used to draw conclusions about the ground-state structures of the GCM for a given dimension.  



\end{abstract}

\maketitle

\section{Introduction}

The statistical mechanics of many-particles systems play a central role in the design and control of materials.  The thermodynamic and transport properties of a material system are determined by the interactions among constituent particles, and these interactions are intimately related to the short- and long-range order in the system.  Although many-body interactions certainly influence the determination of physical properties in these systems, one may generally obtain an accurate approximation to the underlying physics by considering pair interactions $\phi(r), r = \lVert \mathbf{x} - \mathbf{x}^{\prime}\rVert$, between particles. Systems of particles interacting via either inverse-power-laws (e.g., $\phi(r) \sim 1/r^n, n \in \mathbb{N}$) \cite{We81, LiMlGoKa07} or hard core ($\phi(r) = +\infty, 0\leq r<D; \phi(r) = 0, r>D$) \cite{ReFrLe59, MeSi68, FrRiWy85} pair potentials have been extensively studied in this regard.  

For interacting polymers, however, each of the spatially-extended macromolecules has a significant number of degrees of freedom.  Modeling the pairwise monomer interactions in this system can rapidly become computationally prohibitive.  Fortunately, the problem may be simplified immensely by considering instead the interactions among the centers-of-mass of the polymers.  This assumption is equivalent to enforcing an \emph{effective} interaction among the macromolecules.  However, since it is certainly possible for the centers-of-mass of any two polymers to overlap, this effective potential must contain the the essential property of being \emph{bounded}.  The Flory-Krigbaum pair potential $\phi_{\text{FK}}(r)$, introduced in 1950, provides the following form for the effective interaction between the centers-of-mass of two polymer chains:\cite{FlKr50}
\begin{equation}\label{intro1}
\beta \phi_{\text{FK}}(r) = N^2 \left(\frac{V_{\text{seg}}}{V_{\text{solv}}}\right)\left(\frac{3}{4\pi R_g^2}\right)^{3/2} (1-2\chi) \exp\left[\frac{-3r^2}{4 R_g^2}\right],
\end{equation}
where $V_{\text{seg}}$ and $V_{\text{solv}}$ denote the volumes of a monomer segment and a solvent molecule, respectively, $N$ is the degree of polymerization, $R_g$ is the radius of gyration of the chains, $\chi$ is a parameter that controls the solvent quality [$0 < \chi < 1/2$ denotes a good (i.e., conducive to repulsion) solvent and $\chi > 1/2$ a poor one (conducive to attraction)], and $\beta = 1/(k_B T)$ denotes the reciprocal temperature scaled by Boltzmann's constant $k_B$.

The form of \eqref{intro1} suggests that we consider as a general form of the effective pair potential for this system:
\begin{equation}\label{intro2}
\phi(r) = \epsilon \exp\left[-\left(\frac{r}{\sigma}\right)^2\right],
\end{equation}
which is the pair interaction for the so-called Gaussian core model (GCM), originally introduced by Stillinger.\cite{St76}  Here, $\epsilon$ and $\sigma$ determine the energy and length scales, respectively, for the system.  The physical properties of this model have been well-documented up to three dimensions; it is known that the system may undergo a fluid-solid phase transition for sufficiently low temperatures ($k_B T/\epsilon \sim 0.01$), and within the solid-phase region there exists a FCC-BCC ($d = 3$) transition as the system passes from low density to high density.\cite{LaLiWaLoe00}  Furthermore, the GCM displays re-entrant melting, in which the melting temperature $T_m (\rho)$ as a function of density $\rho~(= N/V)$ approaches 0 in the limit $\rho \rightarrow +\infty$; in other words, a crystal in the GCM at positive temperature can always be made to melt by isothermal compression.\cite{St76}  Similarly, $T_m(\rho) \rightarrow 0$ as $\rho \rightarrow 0$; this behavior follows directly from the reduction of the GCM to a system of hard spheres in this limit.\cite{St76} 

Despite the extent of research currently being pursued in this field, little is known about the high-dimensional properties of the GCM.  This gap in knowledge is in spite of the recent interest in the physics of high-dimensional systems of particles; for example, Torquato and Stillinger have previously examined the question of packing hard spheres in high dimensions.\cite{ToSt06, ToUcSt06, SkDoStTo06}  This problem has applications to abstract algebra, number theory, and communications theory, where the optimal method of sending digital signals over noisy channels corresponds to the densest sphere packing in a high-dimensional space.\cite{CoSl}  Besides providing an improvement on the Minkowski lower bound on the maximal packing density in $d$-dimensional Euclidean space $\mathbb{R}^d$, Torquato and Stillinger were also able to provide evidence for a decorrelation principle of disordered packings in high dimensions.\cite{ToSt06}  This principle states that as the dimension $d$ increases, all \textit{unconstrained} correlations vanish, and any higher-order correlation functions $g^{(n)} (\mathbf{x}_1, \dotsc, \mathbf{x}_n)$ may be written in terms of the number density $\rho$ and the radial distribution function $g_2 (r)$ within some small error.  For equilibrium systems, this simplification implies that $g_2 (r)$ approaches its low-density limit as its dimensional asymptotic limit; i.e., the high-dimensional behavior of $g_2 (r)$ is similar to its low-density behavior for any finite $d$.  

With regard to classical fluids, Frisch and Percus \cite{FrPe99} have examined the Mayer cluster expansions for a pair-interacting system in high dimensions and have shown that for repulsive interactions, the series are dominated by ring diagrams at each order in particle density $\rho$.  Resummation of the series leads to an analytic extension in density from which the second virial truncation remains valid at densities higher than the density at which the series diverge.  Doren and Herschbach \cite{DoHe86} previously have developed a dimensionally-dependent perturbation theory for quantum mechanical systems from which they draw conclusions about the energy eigenvalues in ``physical'' dimensions from the information obtained for values of $d$ where simplifications in the behavior of the systems may occur.  It is thus clear that one may obtain keen insight into the physical nature of a many-body system from an exploration of its high-dimensional analogs.

As a result, our focus in the present study is on the phase properties of the GCM in arbitrary Euclidean dimension $d$.  We make the preliminary disclaimer that when we henceforth speak of arbitrary dimension $d$, we imply a Euclidean geometry ($d \in \mathbb{N}$).  It is significant to note that the GCM is ideal for this analytical study since it has the property that $\phi \in L^p (\mathbb{R}^d)~\forall p \in [1, \infty]$.  Therefore, $\phi$ is absolutely integrable, and the Fourier and inverse Fourier transforms are uniquely defined and constitute an isometry.\cite{LiLo}  We utilize the following definition of the Fourier transform (FT) of a function $f(\mathbf{x})$:
\begin{equation}\label{intro2a}
\hat{f}(\mathbf{k}) = \int_{\mathbb{R}^d} \exp\left[-i (\mathbf{k}, \mathbf{x})\right] f(\mathbf{x}) d\mathbf{x},
\end{equation}
where $\hat{f}$ denotes the FT of $f$ and $(\mathbf{k}, \mathbf{x}) = \sum_{i=1}^d k_i x_i$ denotes the inner product of two (real-valued) $d$-dimensional vectors.  Similarly, the inverse FT is defined as:
\begin{equation}\label{intro2b}
f(\mathbf{x}) = \left(\frac{1}{2\pi}\right)^d \int_{\mathbf{R}^d} \exp\left[i (\mathbf{k}, \mathbf{x})\right] \hat{f}(\mathbf{k}) d\mathbf{k}.
\end{equation}
For radial functions [i.e., $f(\mathbf{x}) = f(\lVert\mathbf{x}\rVert)= f(r)$], \eqref{intro2a} and \eqref{intro2b} take the form:\cite{Sn}
\begin{align}
\hat{f}(k) &= (2\pi)^{d/2}\int_0^{+\infty} r^{d-1} f(r) \frac{J_{(d/2)-1}(kr)}{(kr)^{(d/2)-1}} dr\label{intro2c}\\
f(r) &= \left(\frac{1}{2\pi}\right)^{d/2}\int_{0}^{+\infty} k^{d-1} f(k) \frac{J_{(d/2)-1}(kr)}{(kr)^{(d/2)-1}} dk\label{intro2d}.
\end{align}

The FT $\hat{\phi}(k)$ of the pair potential $\phi(r)$ in the GCM is given for all $d$ by:
\begin{equation}\label{intro3}
\hat{\phi}(k) = (\pi \sigma^2)^{d/2} \epsilon \exp\left[-\frac{(k\sigma)^2}{4}\right],
\end{equation}
and the integral of $\phi$ over $\mathbb{R}^d$ is (for $\mathbf{x}, \mathbf{x}^{\prime} \in \mathbb{R}^d$):
\begin{equation}\label{intro4}
\int_{\mathbb{R}^d} \phi(\lVert \mathbf{x}-\mathbf{x}^{\prime}\rVert) d\mathbf{x} = \epsilon (\pi \sigma^2)^{d/2}.
\end{equation}
We immediately see that the dimensionality of the problem is contained entirely in the factors $(\pi \sigma^2)^{d/2}$, facilitating the generalization of the model to arbitrary dimensionality.  Our goal is to characterize the fluid and solid phases of the GCM in high dimensions.  

With regard to the fluid phase, Stillinger and coworkers have developed high-temperature expansions for the excess free energy \cite{St79} $f(\beta)$ and radial distribution function \cite{RoStWa88} $g_2 (r)$, the former for arbitrary $d$ and the latter for $d = 3$.  The convergence properties of the free energy expansion have been previously explored;\cite{St79} although the series for $f(\beta)$ is divergent, it may be formally evaluated via use of Borel resummation.  However, similar convergence properties for the $g_2 (r)$ series remain unestablished even for $d = 3$. Furthermore, while comparisons in three dimensions have been drawn between ``exact'' representations of $g_2 (r)$, either from molecular simulations \cite{LoBoHa00, LaLiWaLoe00} or the aforementioned expansions,\cite{RoStWa88} and numerical approximations, little is known of the high-dimensional applicability of numerical methods.  Of particular interest is the validity of the ``mean-field approximation'' (MFA) to the direct correlation function $c(r)$ that arises from a density functional description of the GCM.\cite{LoBoHa00, LiMlGoKa07, LaLiWaLoe00}  By extending the temperature expansion of $g_2 (r)$ and generalizing to arbitrary dimension, we attempt to elucidate the relationship among the MFA, hypernetted-chain (HNC) approximation, and the Percus-Yevick (PY) approximation to the GCM  and relate the results to a decorrelation principle.  Simultaneously, we explore the convergence properties of the high-temperature expansion of $g_2 (r)$ in arbitrary dimension and relate the results to the phase behavior of the model.  

Prior work on the solid phase of the GCM involving lattice summation energies calculated in $d = 3$ for simple cubic (SC), body-centered cubic (BCC), face-centered cubic (FCC), hexagonal close-packed (HCP), and diamond (DIA) structures indicates a transition in the minimum energy of the system at $\rho \approx \pi^{-3/2}$ from FCC to BCC.\cite{St76}  This conclusion has been supported and expanded to $d = 1, 2$ via the calculation of duality relationships relating the energy per particle $\left(\Phi/N\right)_\Lambda$ at low density of a lattice $\Lambda$ to the corresponding $\left(\Phi/N\right)_{\Lambda^*}$ of the dual lattice $\Lambda^*$.\cite{St79b, StSt97}  However, there remains an open question of the relative stability of lattices in higher dimensions with respect to minimization of the lattice summation energy.  This problem becomes especially apparent for $d \geq 6$, where the family of lattices $D_d$ to which FCC belongs no longer represents the densest known sphere packing among lattices.\cite{CoSl}  Worthy of mention in this regard is the corresponding conjecture by Torquato and Stillinger\cite{ToSt07} that the Gaussian core potential and any other sufficiently well-behaved completely monotonic potential function share the same ground-state structures in $\mathbb{R}^d$ for $2 \leq d \leq 8$ and $d = 24$ although not necessarily at the same densities; more specifically, they claim that these ground states are the Bravais lattices corresponding to the densest known sphere packings for $0 \leq \rho \leq \rho_1$ and the corresponding reciprocal Bravais lattices for $\rho_2 \leq \rho < +\infty$, where $\rho_1$ and $\rho_2$ are the density limits for the phase coexistence region of the lattices.\cite{ToSt07}  We seek to provide numerical support for the latter part of this conjecture with respect to the GCM. 

To these ends, we begin in Section II by developing the requisite high-temperature cluster expansion in $\beta$ for $\ln [g_2 (r)]$ and explore the convergence properties of the series for arbitrary $d$.  At low densities for which the series is appropriate, information about the behavior of the dilute fluid regime of the GCM may be thus obtained.  In Section III we explore the dense fluid regime of the GCM using the three numerical approximations listed above in order to obtain information about $g_2 (r)$ and the associated structure factor $S(k)$ for the system.  The validity of these approximations for arbitrary $d$ is then evaluated.  It may be shown that $S(k)$ approaches a step function with discontinuity at $k = +\infty$ in the infinite-dimensional limit; we use this information to provide analytical support for a decorrelation principle in the fluid phase of the GCM.  We devote Section IV to the solid phase of the GCM.  The behavior of the melting temperature $T_m (\rho)$ in the limit $\rho \rightarrow +\infty$ is generalized with respect to $d$, providing evidence for a fluid-solid phase transition at high density and sufficiently low temperature.  We then calculate lattice summation energies for the $A_d, D_d, E_d,\text{and}~\mathbb{Z}^d$ lattice families and their duals in various dimensions.  After establishing duality relationships for these lattice families, we explore the phase coexistence regions between the lowest-energy lattices and their duals via Maxwell double-tangent constructions and show that the width of these regions increases with respect to the self-dual density $\bar{\rho}^*$ as the dimensionality increases.  The information gathered from this analysis provides evidence for the Torquato-Stillinger conjecture\cite{ToSt07} mentioned above concerning the ground states of certain classical many-particle systems.  Concluding remarks are given in Section V.

\section{Dilute fluid-phase virial behavior of the GCM}

\subsection{Correlation function formalism}

Although there is no evidence for a conventional gas-liquid phase transition in the GCM, it is mathematically convenient to consider the ``dilute'' and ``dense'' fluid regimes separately; the reasons for this distinction will become clear momentarily.  Our understanding of the dilute (i.e., low density) fluid phase will involve the analytical determination of the radial distribution function $g_2 (r)$ in terms of an infinite series in $\beta$.  To motivate this correlation function, we recall that for a system of $N$ (fixed) particles, the configurational part of the canonical partition function is given by:
\begin{equation}\label{dilute1}
Z_N = \idotsint_{V^N} \exp\left[-\beta \Phi(\mathbf{x}_1, \dotsc, \mathbf{x}_N)\right] d\mathbf{x}_1 \dotsm d\mathbf{x}_N,
\end{equation}
where $\Phi(\mathbf{x}_1, \dotsc, \mathbf{x}_N) = \sum\limits_{1\leq i<j\leq N} \phi(r_{ij})$ denotes the total potential energy in the system.  As a result, the probability distribution for observing the many-body system with configuration $x^N = \{\mathbf{x}_1, \dotsc, \mathbf{x}_N\}$ is:
\begin{equation}\label{dilute2}
P(x^N) = \exp\left[-\beta\Phi(x^N)\right]/Z_N.
\end{equation}
Based on \eqref{dilute2}, we define the $n$-body correlation function $\rho^{(n; N)}(\mathbf{x}_1, \dotsc, \mathbf{x}_n)$ to be the joint probability distribution function that in an $N$-particle system, particles will be found at positions $\{\mathbf{x}_1, \dotsc, \mathbf{x}_n\}$.  Mathematically,
\begin{equation}\label{dilute3}
\rho^{(n; N)}(\mathbf{x}_1, \dotsc, \mathbf{x}_n) =  \frac{N!}{(N-n)!} \idotsint_{V^{(N-n)}} P(x^N)~d\mathbf{x}_{n+1}\dotsm \mathbf{x}_N,
\end{equation}
where the prefactor $N!/(N-n)!$ denotes the number of ways of choosing an ordered subset of $n$ particles from a total population of $N$.  For an isotropic fluid (as found in the GCM), we have the result that:
\begin{equation}\label{dilute4}
\rho^{(1; N)}(\mathbf{x}_1) = N/V = \rho.
\end{equation}
It is therefore reasonable to introduce the functions $g_n(\mathbf{x}_1, \dotsc, \mathbf{x}_n)$, defined by:
\begin{equation}\label{dilute6}
g^{(n)}(\mathbf{x}_1, \dotsc, \mathbf{x}_n) = \rho^{(n; N)}(\mathbf{x}_1, \dotsc, \mathbf{x}_n)/\rho^n.
\end{equation}

It is significant to note from the formulation in \eqref{dilute6} that when $g^{(n)}(\mathbf{x}_1, \dotsc, \mathbf{x}_n) = 1$, there is an absence of correlations in the system.  Our primary concern in this study is with the function $g^{(2)}(\mathbf{x}_1, \mathbf{x}_2)$, which for an isotropic fluid takes the form $g_2 (r)$ with $r = \lVert\mathbf{x}_1 - \mathbf{x}_2\rVert$.  The function $g_2$ is known as the \emph{radial distribution function} or \emph{pair correlation function}; since
\begin{equation}\label{dilute7}
\frac{\rho^{(2)}(0, \mathbf{x})}{\rho} = \rho g_2 (r),
\end{equation}
we identify $g_2$ as being proportional to the conditional probability density that a particle will be found at radial distance $r$ given that another is at the origin.  Equivalently, $\rho g_2(r)$ is the average particle density at radial separation $r$ given that a particle is located at the origin.

Since for a general fluid the correlations in the system will diminish with increasing radial separation $r$, we have the asymptotic behavior $g_2 (r) \rightarrow 1$ as $r \rightarrow +\infty$.  It is conventional also to introduce the so-called \emph{total correlation function} $h(r)$, defined by:
\begin{equation}\label{dilute8}
h(r) = g_2(r) - 1.
\end{equation}
It follows from the properties of $g_2$ that $h(r) \rightarrow 0$ as $r \rightarrow +\infty$.


\subsection{Mayer cluster expansion of $g_2$}

It has been well-documented that the radial distribution function $g_2$ may be expanded as an infinite series in the density $\rho$; this expansion has the form:\cite{Rue99}
\begin{equation}\label{dilute9}
g_2 (r) = \exp\left[-\beta\phi_{1,2}\right]\left\{1+\sum_{m=1}^{+\infty}\frac{\rho^m}{m!}\int\left[\sum_{G_{m+2}}\left(\prod_{\alpha \in G_{m+2}} f_{\alpha}\right)\right] \prod_{i=3}^{m+2} d\mathbf{x}_i\right\},
\end{equation}
where $G_{m+2}$ denotes the graph containing $m$ integrable vertices and two stationary vertices which becomes biconnected when an edge is added between the two stationary vertices.  The parameter $\alpha = (i,j)$ corresponds to an edge in the graph $G_{m+2}$, and $f_{\alpha} = \exp\left[-\beta\phi_{\alpha}\right]-1$ denotes the Mayer $f$-function with $\phi_{\alpha}$ representing the pair potential governing the interaction between particles located at $\mathbf{x}_i$ and $\mathbf{x}_j$.  A \emph{biconnected} graph is a collection of vertices and corresponding edges such that one may trace a path between any two vertices even upon the removal of an edge.  We define an \emph{integrable} vertex to be any vertex in the graph $G$ with a corresponding variable of integration $\mathbf{x}_i$ in \eqref{dilute9}; a \emph{stationary} vertex has no corresponding variable of integration.  Unless otherwise stated, vertices 1 and 2 will always be defined as the stationary vertices of any graph $G_{m+2}$, meaning that $g_2$ remains a function of $r = \lVert \mathbf{x}_1 - \mathbf{x}_2\rVert$.  For example, the expansion of $g_2$ up to $O(\rho^2)$ is:
\begin{equation}\label{dilute10}
g_2 (r) = \exp\left[-\beta\phi_{1,2}\right]\left[1+\rho\int f_{1,3} f_{2,3} d\mathbf{x}_3 + O(\rho^2)\right].
\end{equation}
We note that in the limit $\rho \rightarrow 0$, $g_2 (r) \rightarrow \exp\left[-\beta \phi(r)\right],$ which is the Boltzmann factor for the pair potential $\phi$.  For the GCM, it is more useful to pass to an expansion of $g_2$ in reciprocal temperature $\beta$; this conversion is accomplished by a Taylor expansion of each of the $f_{\alpha}$ as follows:
\begin{align}
f_{\alpha} &= \exp\left[-\beta \phi_{\alpha}\right] - 1\label{dilute11}\\
&= \beta \phi_{\alpha} + \left(\frac{1}{2}\right)\left(-\beta \phi_{\alpha}\right)^2 + \dotsb\label{dilute12}.
\end{align}
Subsequent multiplication of the $f$-functions and recollection of orders of $\beta$ leads to the desired expansion.  The reason for passing to the reciprocal temperature expansion is that each of the integrals in \eqref{dilute9} reduces to an integral over products of Gaussians, which may usually be evaluated analytically by repeated use of the relation (for $\epsilon = \sigma = 1$):
\begin{equation}\label{dilute13}
\int_{\mathbb{R}^d} \phi_{i, j}^m \phi_{j, k}^n d\mathbf{x}_j = \left(\frac{\pi}{m+n}\right)^{d/2} \phi_{i, k}^{\frac{mn}{m+n}}.
\end{equation}
Equation \eqref{dilute13} follows directly from \eqref{intro4}.

Although, as mentioned above, it is possible to analytically evaluate each of the terms in the $\beta$ series  derived from \eqref{dilute9}, the mathematical complexity of the problem increases significantly around $O(\beta^6)$ due to the increasing number of integrals to evaluate.  However, it turns out that several of the graphs from this expansion may be ``removed'' by passing to the expansion of $\ln\left[g_2(r)\right]$, which may be suitably derived from \eqref{dilute9} by a Taylor expansion.  The advantages of the logarithmic expansion are that (a) it contains the entire $\exp\left[-\beta \phi_{1,2}\right]$ term from \eqref{dilute9} in $-\beta \phi_{1,2}$ and that (b) it removes all parallel graphs from the expansion, thereby drastically reducing the computational cost of analytical evaluation.  In general, this series takes the form:
\begin{equation}\label{dilute16}
\ln\left[g_2(r)\right] = \sum_{n=1}^{+\infty} \left(-\beta\right)^n f_n (r).
\end{equation}

An investigation of convergence properties of the $\beta$-series in \eqref{dilute16} may be done via reference to the ratio test for infinite series.\cite{Ru64}  We consider the ratios defined by:
\begin{equation}\label{dilute17}
\zeta_{n+1} (r) = \left\lvert \frac{f_{n+1}(r)}{f_n (r)} \right\rvert = \frac{f_{n+1} (r)}{f_n (r)}.
\end{equation}
The radius of convergence $\lambda$ of the series is then given by:
\begin{equation}\label{dilute18}
\limsup_{n\rightarrow \infty} \zeta_{n+1} (r) = \frac{1}{\lambda}.
\end{equation}
We note that the left-hand side of \eqref{dilute18} is guaranteed to exist $\forall r \in [0, \infty)$ although it may be $\pm \infty$.  It is important to note that the proof of convergence of the density expansion \eqref{dilute9} is well-known and can be found, for example, in Ruelle.\cite{Rue99}  All that is required is to show that the GCM pair interaction is stable and regular.  Borrowing the definitions from Ruelle,\cite{Rue99} a $k$-body interaction $(\phi^{(k)})_{k\geq2}$ is \emph{stable} if there exists a $B \geq 0$ such that:
\begin{equation}\label{dilute19}
\Phi(\mathbf{x}_1, \dotsc, \mathbf{x}_n) \geq -nB
\end{equation}
for all $n \geq 0$ and $\mathbf{x}_1, \dotsc, \mathbf{x}_n \in \mathbb{R}^d$.  A pair interaction $\phi$ is \emph{regular} if it is bounded from below by a finite constant $K$ and satisfies:
\begin{equation}\label{dilute20}
C(\beta) = \int\lvert\exp\left[-\beta\phi(\mathbf{x})\right]-1\rvert d\mathbf{x} < +\infty
\end{equation}
for some $\beta > 0$ and hence for all $\beta > 0$.  The proof that the GCM pair potential is regular can be found in Appendix A. 

\begin{figure}[!htp]
\includegraphics[width=0.7\textwidth]{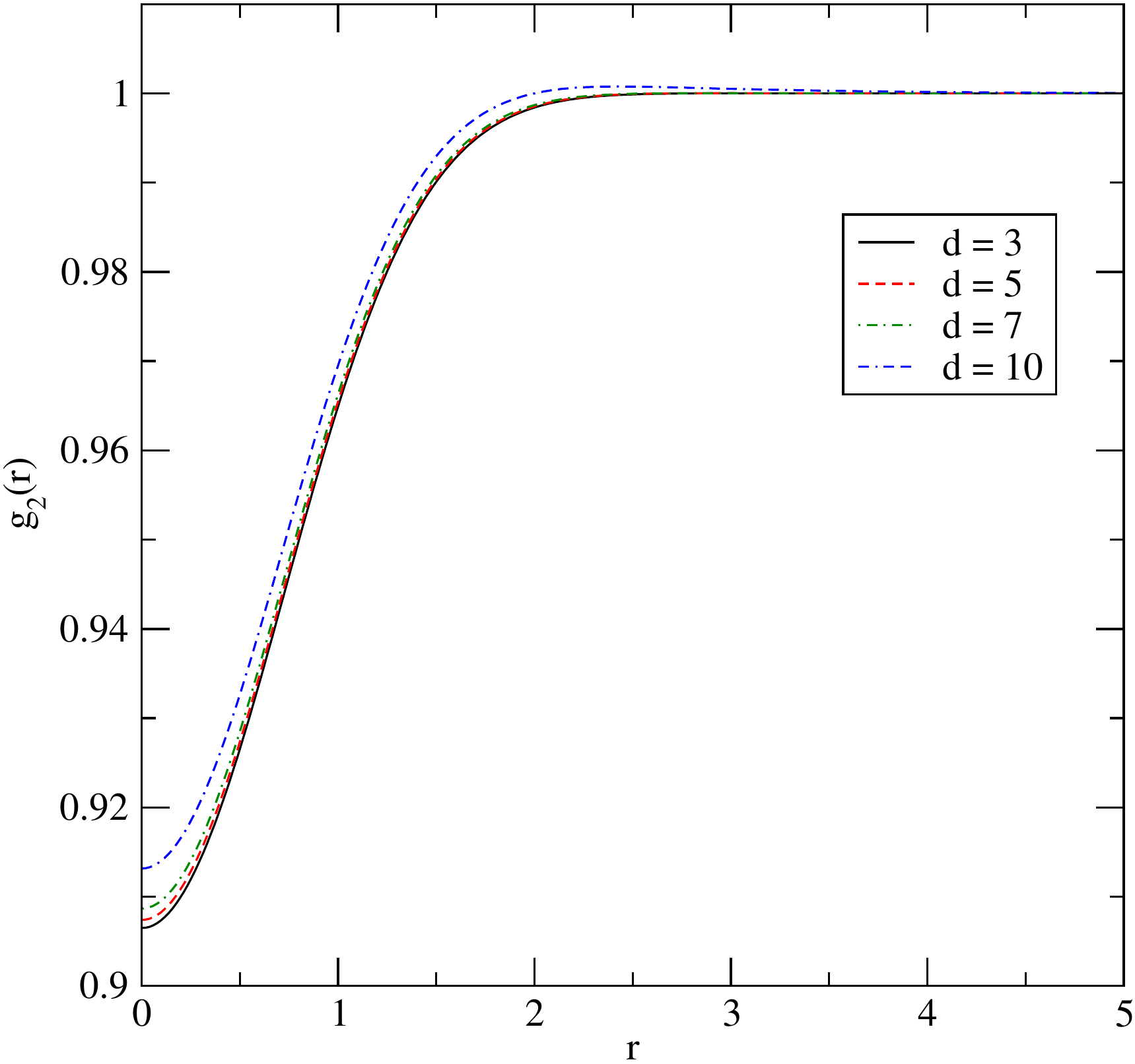}
\caption{\label{figone}The radial distribution function $g_2$ for $\rho = 0.1, \beta = 0.1$ as obtained from the summation in \eqref{dilute16} to $O[(-\beta)^5]$.}
\end{figure}

We have successfully evaluated each of the $f_n(r)$ in \eqref{dilute16} up to $O[(-\beta)^5]$ with the results (in units such that $\epsilon = \sigma = 1$):
\begin{equation}\label{rdilute1}
f_1 (r) = \phi
\end{equation}
\begin{equation}\label{rdilute2}
f_2 (r) = \rho \left(\frac{\pi}{2}\right)^{d/2} \phi^{1/2}
\end{equation}
\begin{equation}\label{rdilute3}
f_3 (r) = \rho \left(\frac{\pi}{3}\right)^{d/2} \phi^{2/3} + \rho^2 \left(\frac{\pi^2}{3}\right)^{d/2} \phi^{1/3}
\end{equation}
\begin{equation}\label{rdilute4}
\begin{split}
f_4 (r) &= \rho \left(\frac{1}{3}\right) \left(\frac{\pi}{4}\right)^{d/2} \phi^{3/4} + \rho \left(\frac{1}{4}\right) \left(\frac{\pi}{4}\right)^{d/2} \phi + \rho^2 \left(\frac{3}{2}\right) \left(\frac{\pi^2}{5}\right)^{d/2} \phi^{2/5}\\
&+ 2 \rho^2 \left(\frac{\pi^2}{5}\right)^{d/2} \phi^{3/5} + \rho^3 \left(\frac{\pi^3}{4}\right)^{d/2} \phi^{1/4}
\end{split}
\end{equation}
\begin{equation}\label{rdilute5}
\begin{split}
f_5(r) &= \rho \left(\frac{1}{12}\right) \left(\frac{\pi}{5}\right)^{d/2} \phi^{4/5} + \rho \left(\frac{1}{6}\right) \left(\frac{\pi}{5}\right)^{d/2} \phi^{6/5} + \rho^2 \left(\frac{1}{2}\right) \left(\frac{\pi^2}{7}\right)^{d/2} \phi^{3/7}\\ 
&+ \rho^2 \left(\frac{3}{4}\right) \left(\frac{\pi^2}{8}\right)^{d/2} \phi^{1/2} + \rho^2 \left(\frac{\pi^2}{7}\right)^{d/2} \phi^{5/7} + \rho^2 \left(\frac{\pi^2}{7}\right)^{d/2} \phi^{6/7}\\ 
&+ 2 \rho^2 \left(\frac{\pi^2}{8}\right)^{d/2} \phi^{5/8} + 2 \rho^3 \left(\frac{\pi^3}{7}\right)^{d/2} \phi^{2/7} + \rho^2 \left(\frac{1}{2}\right) \left(\frac{\pi^2}{8}\right)^{d/2} \phi\\ 
&+  3\rho^3 \left(\frac{\pi^3}{8}\right)^{d/2} \phi^{3/8} + 2 \rho^3 \left(\frac{\pi^3}{7}\right)^{d/2} \phi^{4/7} + \rho^3 \left(\frac{\pi^3}{8}\right)^{d/2} \phi^{1/2}\\
& + \rho^4 \left(\frac{\pi^4}{5}\right)^{d/2} \phi^{1/5},
\end{split}
\end{equation}
where, as above, $\phi = \phi(r_{12})$ denotes the pair interaction between two particles in the GCM.  Using the results in \eqref{rdilute1}-\eqref{rdilute5}, we may obtain an approximate plot of $g_2$ in the limit of high temperature and low density, indicative of a dilute fluid.  Such a plot is included in Figure \ref{figone} for various values of $d$.

Note that the value of $g_2(0)$ is nonzero in each case; we expect this result since the bounded nature of the potential allows for a finite probability of particle overlap.  Furthermore, the fact that the potential is repulsive suggests the presence of negative correlations near the origin, which we also observe.  Correlations in the dilute fluid phase rapidly diminish as $r$ increases; what is perhaps most significant about $g_2$ in this case, however, is that the correlations also diminish as the dimensionality of the system increases; i.e., $g_2 \rightarrow 1$ more quickly as $d$ increases.  This observation provides the first suggestion (but of course does not prove) that a decorrelation principle applies for the GCM.  We will return to this point in the study of the dense fluid regime.   

To help us understand the high-dimensional behavior of $g_2$, we examine more closely the expansion in \eqref{dilute16}.  Without loss of generality, we will momentarily work with reduced units such that $\sigma = \epsilon = 1$.  Let $C_{n+1}$ denote the unique chain diagram contribution (containing $n+1$ total vertices) to the order $\beta^n$ factor in the $g_2$ high-temperature expansion; namely,
\begin{equation}\label{rdilute6}
C_{n+1} = \rho^{n-1} \int \prod_{i=1}^{n} \phi_{i, i+1} \prod_{i=2}^{n} d\mathbf{x}_i,
\end{equation}
where for notational convenience we have chosen vertices $1$ and $n+1$ as the stationary vertices in the graph.
We show in Appendix B that:
\begin{equation}\label{rdilute7}
C_{n+1} = \rho^{n-1}\left(\frac{\pi^{n-1}}{n}\right)^{d/2} \phi^{1/n}
\end{equation}
$\forall n \geq 2$.  

The argument used to prove \eqref{rdilute7} elucidates a central property of the cluster integrals:  the exponential order of $\pi$ is determined solely by the number of integrable vertices in the corresponding cluster diagram.  This statement is not exclusive to the chain diagrams.  Since the dimensionality $d$ is contained only in the factors $\left(\pi^{\alpha}/\gamma\right)^{d/2}$ that appear for each cluster integral, we see that those diagrams that maximize the value of $\alpha$ for $\gamma$ of order $n$ will dominate the $g_2$ cluster expansion.  Note that for $n \in \mathbb{N}\setminus\{1\}$, $\pi^{n-1} > n$, which may be easily proved by mathematical induction.  Since the chain diagram for the $\beta^n$ contribution to the cluster expansion contains the greatest number of integrable vertices, we expect this diagram to maximize the quotient $(\pi^{\alpha}/\gamma)$ and thus dominate the expansion at any order $\beta^n$ for $\gamma$ of order $n$.  This conclusion is in agreement with previously reported results for classical pair-interacting repulsive fluids by Frisch and Percus.\cite{FrPe99}  Therefore, as the dimension $d$ increases, we expect that $g_2$ may be represented to a good approximation by the summation over these chain diagrams; we will see momentarily that this truncation of the series corresponds to the MFA and that the result is convergent.  


\begin{figure*}
\includegraphics[width=\columnwidth]{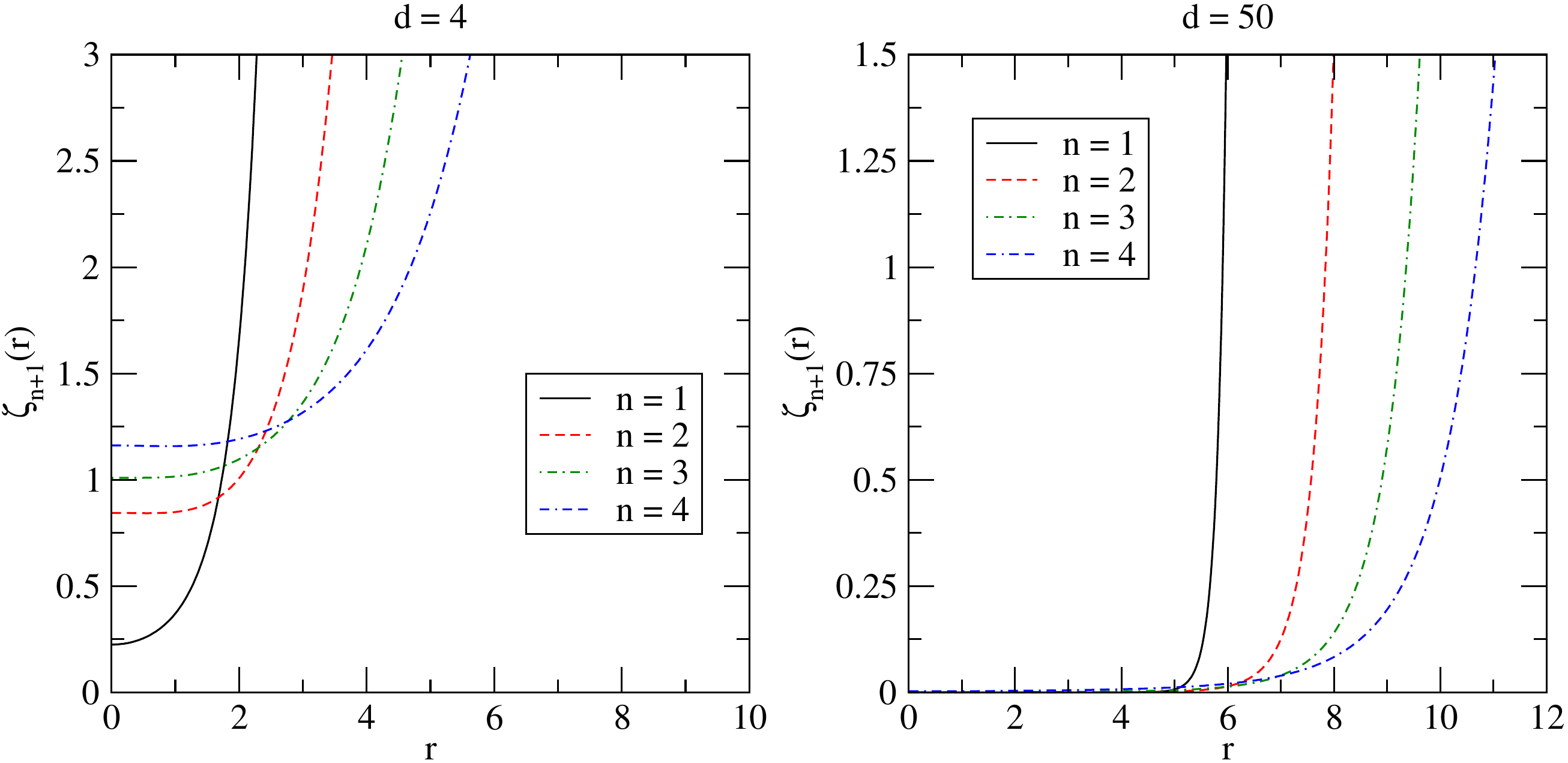}
\caption{\label{figfourd}The ratios $\zeta_{n+1}(r)$ derived from \eqref{rdilute1}-\eqref{rdilute5} for $\bar{\rho} = \rho\pi^{-d/2}$ = 0.9.  Reduced units $\epsilon = \sigma = 1$ have been used.}
\end{figure*}

This convergence claim is \emph{not} true, however, for the general series representation, as may be seen from the plots of the ratios $\zeta_{n+1}(r)$, defined in \eqref{dilute17}, in Figure \ref{figfourd}.  We note that the ratios depend strongly on the dimension $d$ of the system and the radial separation $r$; however, they eventually diverge as $r$ increases.  There is no reason \textit{a priori} why the $g_2$ cluster expansion should converge only for some values of $r$, and we therefore expect the series to diverge as a whole.  We recall that the high-temperature expansion of $g_2$ is centered at $\beta = 0$; therefore, if it is to converge for some $\beta > 0$, it must also converge for certain values of $\beta < 0$.  However, as the reciprocal temperature $\beta$ passes through 0, the pair potential undergoes an effective transition from repulsive to attractive interactions due to the coupling of $\beta$ and the energy scale $\epsilon$ in the Boltzmann factor.  Because the pair potential does not exclude particle overlap, even in the limit $r \rightarrow 0$, the attractive regime forces the particles to cluster on top of each other at a point, and the system undergoes a \emph{collapse instability}.  This collapse instability forces the radius of convergence to $0$ and has been documented by Stillinger for the equivalent high-temperature free-energy expansion.\cite{St79, St80}  In terms of interacting polymers, the collapse instability corresponds to changing the effective composition of the solvent such that aggregation of the macromolecules is energetically favorable.  The FK potential in \eqref{intro1} captures this behavior via variation in the parameter $\chi$.  We make note, however, that the divergence of $\zeta_{n+1}(r)$ appears to push outward toward $r = +\infty$ as $d$ increases.  It is therefore possible that a high-dimensional approximation such as the one mentioned above may be able to overcome the collapse instability, and we explore this possibility in Section III below.     

\section{Dense fluid-phase behavior of the GCM}

The low-density cluster expansion studied in the dilute fluid regime, though convergent in $\rho$ according to Ruelle,\cite{Rue99} has a finite radius of convergence which is necessarily small.  Therefore, to discern information about the fluid phase at values of $\rho$ greater than the radius of convergence of the series above, we rely on approximation methods (numerical and analytical) to estimate $g_2$.  These approximation methods rely on solutions to the so-called \emph{Ornstein-Zernike equation}, given by:
\begin{align}
h(r) &= c(r) + \rho \int_{\mathbb{R}^d} h(\lVert \mathbf{x} - \mathbf{x}^{\prime}\rVert) c(\mathbf{x}^{\prime}) d\mathbf{x}^{\prime}\label{dense1}\\
&= c(r) + \rho(h * c)(r)\label{dense2},
\end{align}  
where $c(r)$ denotes the \emph{direct correlation function}, and $(h*c)$ indicates the convolution of $h$ and $c$.  We take the Ornstein-Zernike (OZ) equation as the definition of the direct correlation function; it essentially separates the immediate interactions between two particles from those interactions that result indirectly from interactions with surrounding particles.  The advantage of the OZ equation is that it allows us to approximate $h$ and therefore $g_2$ by making a reasonable ansatz about the form of $c$.  We consider the following three well-documented closures to the OZ equation (see, e.g., McQuarrie \cite{Mc}):
\begin{align}
c_{\text{HNC}}(r) &= \exp\left[-\beta\phi(r) + \gamma(r)\right] - \gamma(r) -1, \text{ (hypernetted chain)}\label{dense3}\\
c_{\text{PY}}(r) &= \left\{1+\gamma(r)\right\}\left\{\exp\left[-\beta\phi(r)\right]-1\right\}, \text{ (Percus-Yevick)}\label{dense4}\\
c_{\text{MFA}}(r) &= -\beta \phi(r), \text{ (mean-field approximation)}\label{dense5},
\end{align}
where
\begin{equation}\label{dense6}
\gamma(r) = h(r) - c(r).
\end{equation}
Both the PY approximation and the MFA may be obtained from the HNC approximation via linearization of one or more of the exponential functions in \eqref{dense3}.  Solutions to the OZ equation utilizing the HNC and PY approximations are necessarily numerical.  We utilize a relatively simple modified Picard iteration algorithm, a review of which may be found in the article by Busigin and Phillips.\cite{BuPh92}  For reference, we include the details of the algorithm in Appendix C.

The MFA is unique in the sense that one may obtain some analytical results from the OZ equation using Fourier analysis as a result of the relation \eqref{intro3}.  Namely, the OZ equation implies:
\begin{align}
\hat{h}(k) &= \frac{\hat{c}(k)}{1-\rho\hat{c}(k)}\label{dense10}\\
\Rightarrow \hat{h}_{\text{MFA}}(k) &= \frac{-\beta \epsilon \left(\pi\sigma^2\right)^{d/2} \exp\left[-\frac{(\sigma k)^2}{4}\right]}{1+\beta\epsilon\rho(\pi\sigma^2)^{d/2}\exp\left[-\frac{(\sigma k)^2}{4}\right]}\label{dense11}.
\end{align}
Our study of this approximation will involve the associated structure factor factor $S(k)$ for the system at a given density, which we introduce here as:
\begin{equation}\label{dense12}
S(k) = 1 + \rho \hat{h}(k) = \frac{1}{1-\rho\hat{c}(k)},
\end{equation}
where the second equality follows from the OZ equation.  The structure factor is proportional to the scattered intensity of radiation from a system of points and thus is experimentally observable; this notion is the physical motivation for the intrinsic property that $S(k) \geq 0 ~\forall k \in [0, +\infty)$.  Determination of the structure factor for a system is therefore a means to establish the physicality of a given approximation.  For the MFA, note that \eqref{intro3} and \eqref{dense5} imply:
\begin{equation}\label{dense13}
S_{\text{MFA}}(k) = \frac{1}{1+\beta\epsilon\rho(\pi\sigma^2)^{d/2}\exp\left[-\frac{(\sigma k)^2}{4}\right]}.
\end{equation}

\begin{figure*}[!htp]
\includegraphics[width=\columnwidth]{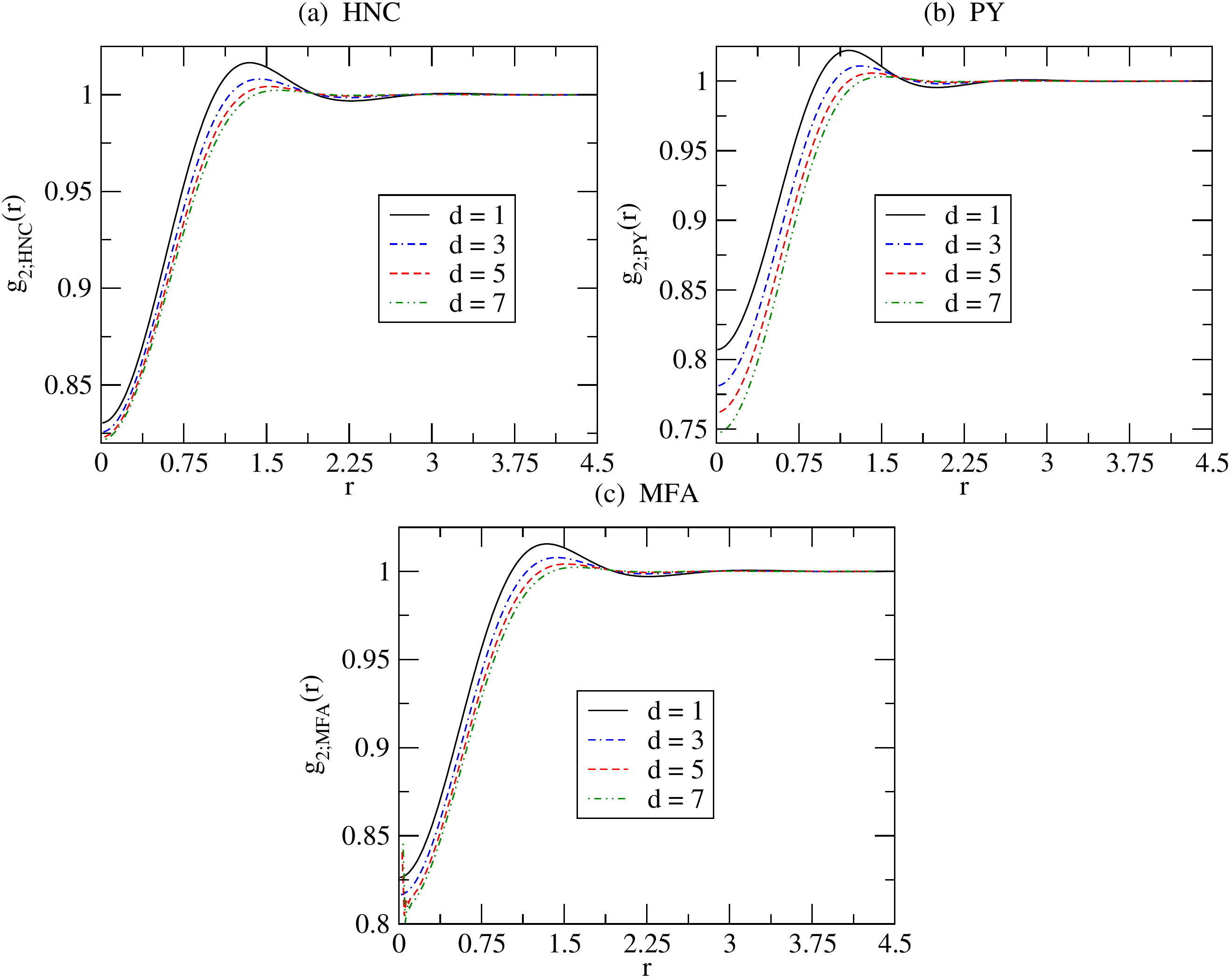}
\caption{\label{figfourb}Approximation methods for the GCM with $d = 3, 5, 7$.  The density $\rho\sigma^d = 5.0$, and $\beta\epsilon = 1.0$.}
\end{figure*}

We collect in Figure \ref{figfourb} the results for $g_2$ as obtained from the HNC and PY approximations along with the MFA utilizing the iterative Fourier algorithm above.  We have chosen the density $\rho$ to be sufficiently high to capture the behavior of the dense fluid regime.  Each of the approximations shows short-range correlations which diminish rapidly with increasing $r$, indicative of gas-like fluid behavior.  Again, the probability of finding particles at zero separation is nonvanishing due to the bounded nature of the potential.  However, while the HNC approximation and the MFA have similar values for $g_2 (0)$, it appears that the PY approximation underestimates this value and thereby introduces an increased effective repulsion among the particles.  Note that the correlations also diminish with dimensionality, providing further numerical support for a decorrelation principle with respect to $g_2$.  


\begin{figure}[!tp]
\includegraphics[width=0.7\textwidth]{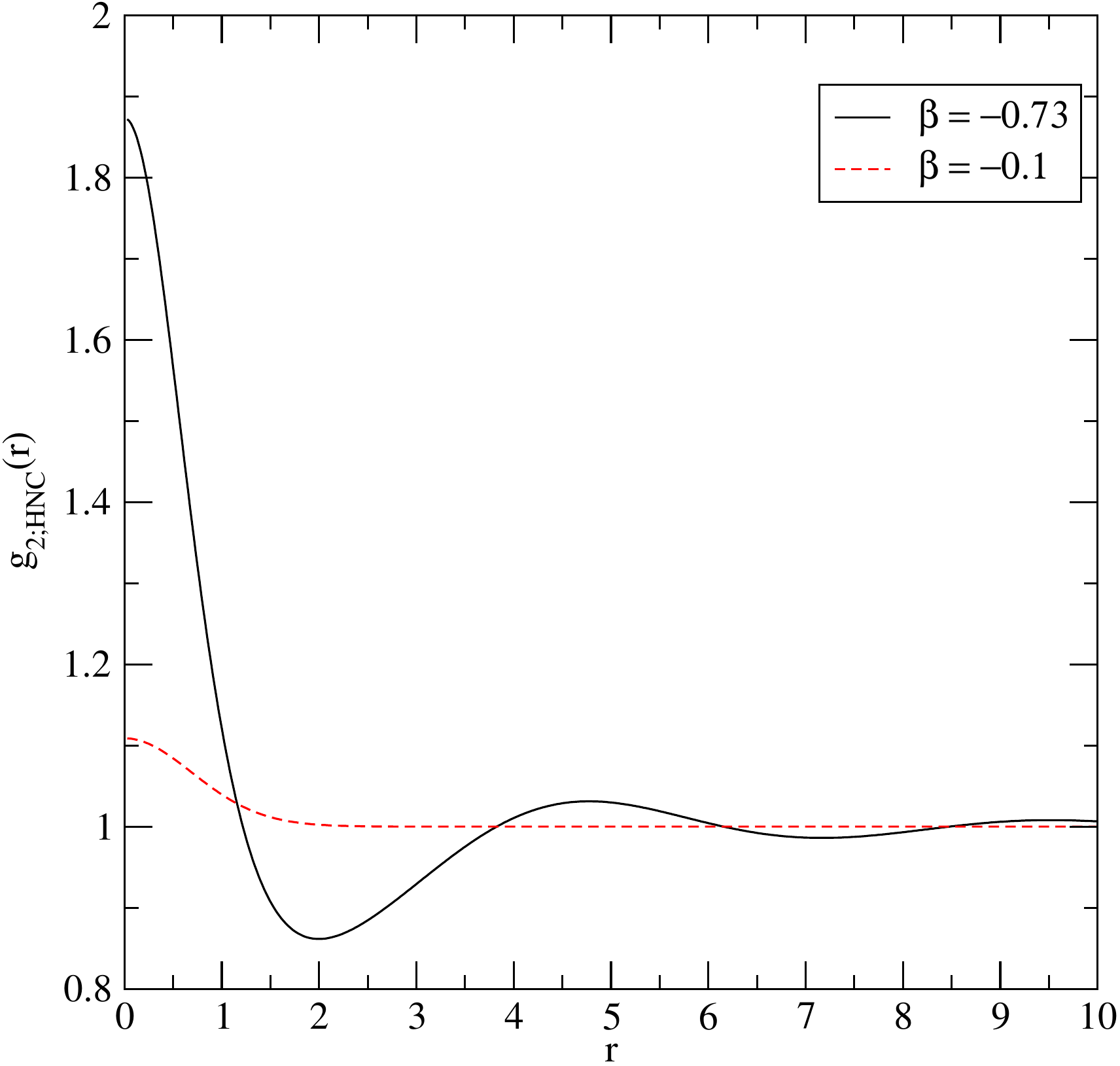}
\caption{\label{figseven}Hypernetted chain approximation to $g_2$ with $d = 5$ demonstrating the (non-divergent) collapse phenomenon for value of $\beta\epsilon < 0$.  The density $\rho\sigma^5 = 0.1$.}
\end{figure}

Figure \ref{figseven} shows the results for the HNC approximation with $d = 5$ and $\beta\epsilon < 0$.  Here, the approximation does not immediately diverge and shows aggregation at the origin; as the value of $\beta\epsilon$ becomes increasingly negative, the correlations in the system become longer in range until the approximation does in fact diverge for $\beta\epsilon$ slightly less than $-0.73$.  Similar results have been priorly reported by Root, Stillinger, and Washington \cite{RoStWa88} for the PY approximation.  Thus, these numerical schemes have the capacity to branch into the instability region for at least some small range of $\beta\epsilon < 0$.  The results obtained in this region certainly do not contain the actual physics of the GCM, but they do reflect the properties of aggregation and increasing correlations among particles that we expect with collapse and thus provide valuable insight into the nature of this instability.        

\begin{figure}[!htp]
\includegraphics[width=0.7\textwidth]{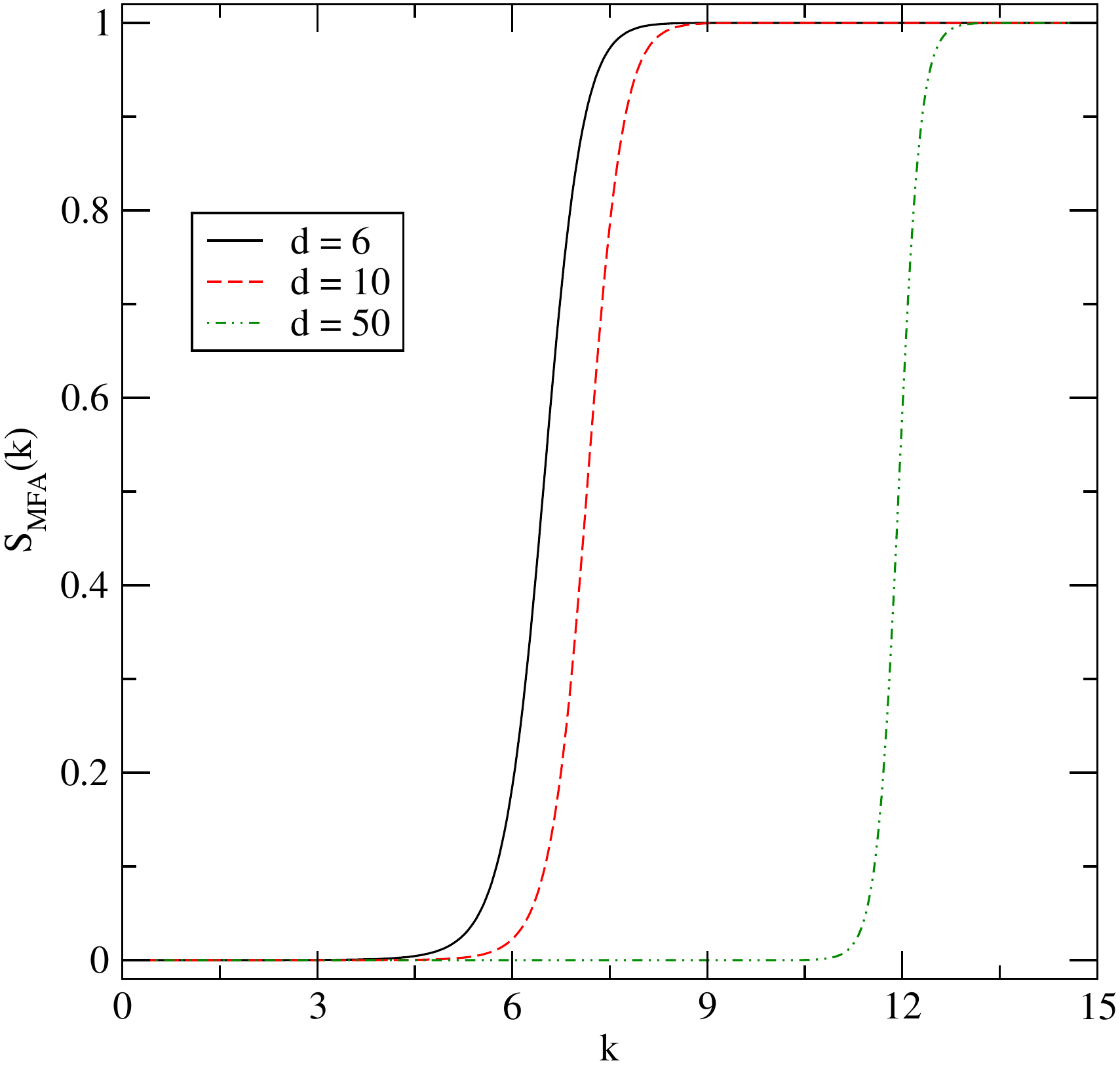}
\caption{\label{figsevenb}Structure factor $S_{\text{MFA}}(k)$ in the MFA for various $d$; $\rho = 36/\pi, \beta = 100$.}
\end{figure}

Let us now turn our attention to some analytical properties of the MFA.  We will for the moment work with unitless parameters such that $\beta^* = \beta\epsilon$, $\rho^* = \rho\sigma^d$, and $k^* = k\sigma$.  For notational convenience, we will continue to denote these quantities as $\beta$, $\rho$, and $k$, respectively.  It is important to note that when we speak of the infinite-dimensional limit, we mean the limit as $d \rightarrow +\infty$ such that $\beta^*$ and $\rho^*$ are held constant.  Define the dimensionally-dependent parameter $\lambda_d = \beta \rho \pi^{d/2}$, which implies (see \eqref{dense11} and \eqref{dense13}):
\begin{equation}
S_{\text{MFA}}(k) = \frac{1}{1+\lambda_d\exp\left[-\frac{k^2}{4}\right]}\label{rdense2}.
\end{equation}
It is clear from \eqref{rdense2} that $S_{\text{MFA}} \rightarrow 1$ as $k\rightarrow +\infty$ for $\lambda_d$ fixed; indeed, one can see from Figure \ref{figsevenb} that $S_{\text{MFA}}$ represents a ``smoothed step function'' for any finite dimension. 

Consider the limit $\lambda_d \rightarrow +\infty$.  Equation \eqref{rdense2} implies that $S_{\text{MFA}} \rightarrow 0$ on any compact subset of $\mathbb{R}^+$ (used here to denote the nonnegative reals) in this case.  We note, however, that this limit does not commute with the limit $k \rightarrow +\infty$; this issue is directly related to the apparent disappearance of the ``step function'' in the high-dimensional limit.  To address this problem, it is worthwhile to consider the evolution of the slope of $S_{\text{MFA}}$ at $S_{\text{MFA}}(k) = 1/2$, the ``midpoint'' of the structure factor, as $\lambda_d \rightarrow +\infty$.  Assuming $\beta \rho \geq 0$ and fixed $d$, $S_{\text{MFA}}(k) = 1/2$ occurs at $k_d = 2 \sqrt{\ln(\lambda_d)} \geq 0$.  The slope of $S_{\text{MFA}}$ at any point $k$ in its domain is given by the derivative $S^{\prime}_{\text{MFA}}$:
\begin{align}
S^{\prime}_{\text{MFA}}(k) &= \frac{\lambda_d k \exp\left[-\frac{k^2}{4}\right]}{2\left(1+\lambda_d\exp\left[-\frac{k^2}{4}\right]\right)^2}\label{rdense3}\\
&= \left[S(k)\right]^2\left(\frac{\lambda_d k}{2}\right)\exp\left[-\frac{k^2}{4}\right]\label{rdense4}.
\end{align}
Evaluating \eqref{rdense4} at $k_d$:
\begin{align}
S^{\prime}_{\text{MFA}}(k)\Bigl\vert_{k = k_d} &= \frac{k_d}{8}\label{rdense5}\\
&= \frac{\sqrt{\ln(\lambda_d)}}{4}\label{rdense6}.
\end{align}
We immediately notice from \eqref{rdense6} that the slope diverges in the infinite-dimensional limit, indicative of the behavior of a true Heaviside step function.  Our claim based on this information is that the MFA approaches a step function with discontinuity at $k = +\infty$ in the infinite-dimensional limit.  We draw the connection here with a system of identical hard spheres, the pair correlation function of which is exactly a Heaviside step function; as $d \rightarrow +\infty$, the mean field approximation to the GCM thereby resembles a system of hard spheres with arbitrarily large radii interacting in the dual space (in the sense of Fourier transforms) to the real space of the Gaussian core particles for nonzero density.   In low-dimensional reciprocal space, the hard spheres are ``smoothed'' by the MFA, meaning interparticle penetration becomes increasing likely with decreasing dimension; however, the particles adopt an increasingly hard core as the dimension increases.  This information allows us to draw some analytical conclusions about a decorrelation principle for the GCM in this regime.  Assuming for sufficiently high dimension we may write as an approximation for $S_{\text{MFA}}$:
\begin{equation}\label{rdense7}
S_{\text{MFA}} = \Theta(k-k_d),
\end{equation}
where $\Theta(k-k_d)$ denotes the Heaviside step function in reciprocal space with discontinuity at $k = k_d$, it is possible to evaluate the coordinate space correlation functions analytically.  The result is:
\begin{align}
h(r) &= -\bar{v}\left(\frac{1}{2\pi}\right)^{d/2} \int_0^{+\infty} k^{d-1} \Theta(k_d-k) \left(\frac{J_{(d/2)-1}(kr)}{(kr)^{(d/2)-1}}\right) dk\label{rdense10}\\
\Rightarrow g_2 (r) &= 1-\bar{v}\left(\frac{1}{2\pi}\right)^{d/2} J_{d/2}(k_d r) \left(\frac{k_d}{r}\right)^{d/2}\label{rdense13},
\end{align}
where $\bar{v} = 1/\rho$ denotes the reciprocal density.
Since we consider $d \gg 1$, we utilize the principal asymptotic form of the Bessel function in \eqref{rdense13} to obtain the result:
\begin{equation}\label{rdense14}
g_2 (r) \sim 1-\bar{v}\left(\frac{e k_d^2}{2}\right)^{d/2} \left(\frac{1}{\pi d}\right)^{(d+1)/2}.
\end{equation}

We make note of the constraint that $\beta \rho \geq \pi^{-d/2}$ to ensure that $k_d \in \mathbb{R}$, restricting this approximation essentially to the low-temperature/high-density regime.  Equation \eqref{rdense14} is single-valued for all $r$; it is clearly less than 1 for any finite dimension, and the value it adopts is roughly the minimum of the expression for $g_2$ in \eqref{rdense13}.  Note that for $d \rightarrow +\infty$ the values of $g_2$ in \eqref{rdense13} and \eqref{rdense14} both approach $g_2(r) = 1$.  The fact that \eqref{rdense13} and \eqref{rdense14} converge in the limit $d \rightarrow +\infty$ thereby reflects a loss of correlations in the infinite-dimensional limit, and from our knowledge of the behavior of the structure factor in the MFA, we expect that this ``hard sphere'' approximation accurately captures the high-dimensional behavior of the system.  These results thereby provide analytical support for a decorrelation principle with the GCM in the fluid phase. 

To examine the behavior of the MFA in the high-temperature limit, we derive the cluster expansion of $g_2^{(\text{MFA})}$ with respect to reciprocal temperature $\beta$.  This is easily accomplished via iteration of the Ornstein-Zernike equation with respect to $h$.  The result is:
\begin{equation}\label{rdense15}
g_2^{(\text{MFA})}(r) = 1-\beta\phi(r) + \sum_{n=2}^{+\infty} (-\beta)^n C_{n+1}(r).
\end{equation}  

\begin{figure}[!htp]
\includegraphics[width=0.7\textwidth]{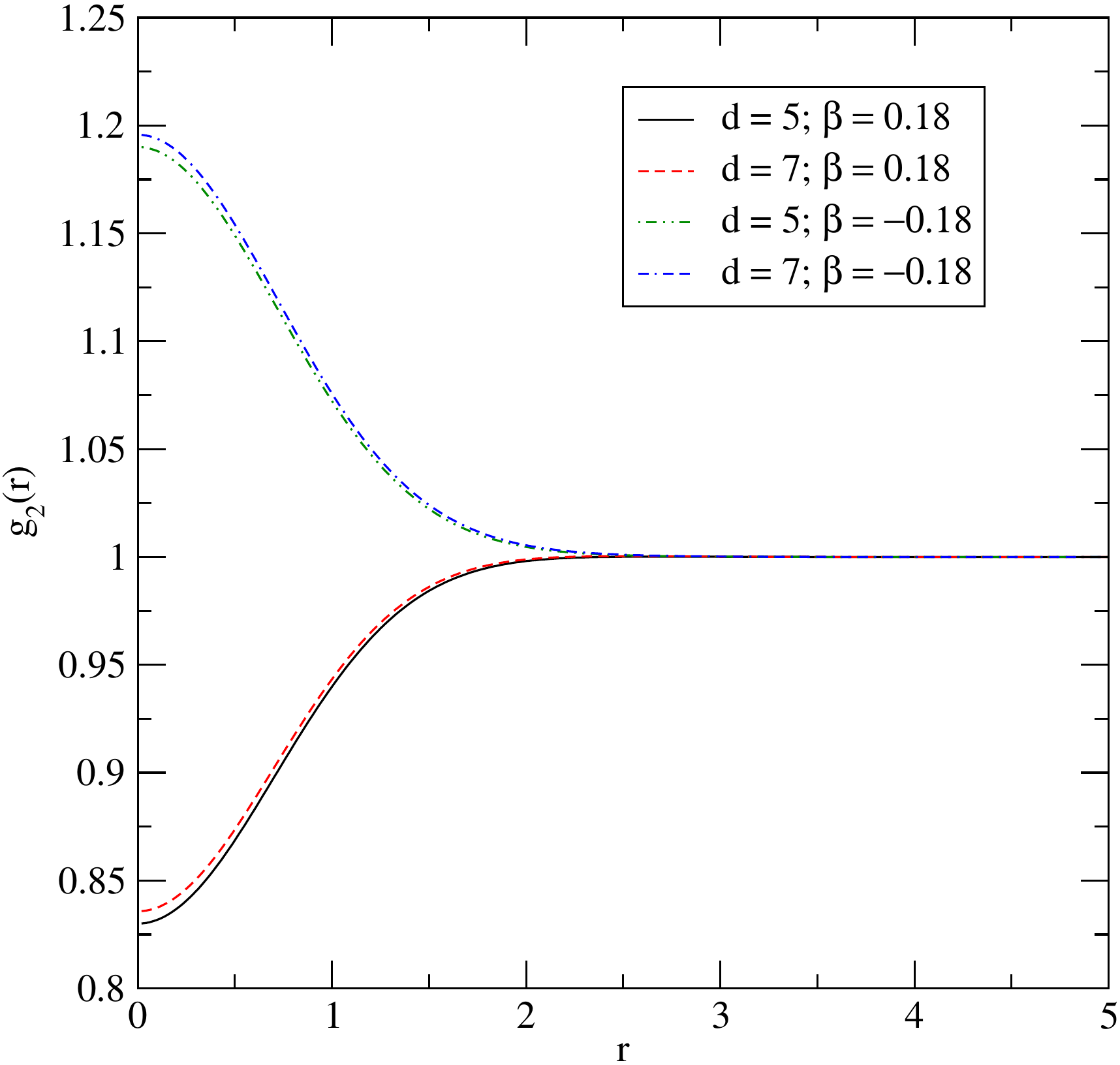}
\caption{\label{fignine}Radial distribution function $g_2$ from the series representation of the MFA in \eqref{rdense15} for $d = 5, 7$ and $\beta = 0.18, -0.18$.  Note that as the MFA passes through $\beta = 0$, the form of $g_2$ reflects the aggregation of the HNC approximation near the collapse instability; however, the series in \eqref{rdense15} is still convergent for some values of $\beta < 0$.}
\end{figure}

The MFA therefore keeps only the dominating terms from the analytical high-temperature expansion.  However, unlike the analytical expansion, the series in \eqref{rdense15} is \emph{convergent}; the proof is left to Appendix D.  Plots of $g_2$ obtained from \eqref{rdense15} are given in Figure \ref{fignine}.  For values of $\beta > 0$ such that the series converges, the form of $g_2$ is very similar to the results obtained from the cluster expansion work above.  In fact, we expect that the MFA becomes a better approximation to the cluster expansion as $d$ increases due to the domination of the series by the chain diagrams.  Since the density expansions for the HNC and PY approximations, which are well-known (see, e.g., McQuarrie \cite{Mc}) and are not derived here, retain the chain diagrams that appear in \eqref{rdense15}, we conclude that these approximation schemes should converge in the high-dimensional limit.  For values of $\beta < 0$ and within the radius of convergence, the MFA expansion shows a similar aggregation phenomenon to the one observed from the HNC approximation; as before, this approximation scheme penetrates into the instability region without collapse up to some finite value of $\beta$, beyond which the series diverges.  We remark that the expression in \eqref{rdense15} provides a computationally convenient way of computing approximations to $g_2$ for the GCM to any order $\beta^n$ since analytic results are available for the chain diagrams.   

\section{Solid-phase behavior of the GCM}

\subsection{High-dimensional lattice structures}

Our focus in the study of the solid phase of the GCM model will primarily involve the determination of lattice summation energies for known lattices up to $d = 8$.  Unfortunately, structural information for lattices in dimensions significantly higher than eight is either unavailable or computationally prohibitive to obtain.  However, the results presented here provide significant evidence in favor of a decorrelation principle.  

A lattice in high dimensions is characterized by the integer-valued linear combinations of a set of primitive basis vectors; i.e., a lattice $\Lambda = \{\mathbf{p}\}$, where:
\begin{equation}\label{solid1}
\mathbf{p} = \sum_{i=1}^{d} n_i \mathbf{a}_i.
\end{equation}
In the expression above $n_i \in \mathbb{N} ~\forall i,$ and $\mathbf{a}_i$ denotes the $i$-th basis vector for the lattice.   Associated with a lattice $\Lambda$ is the so-called dual lattice $\Lambda^*$ with basis vectors $\mathbf{q}$ defined by $(\mathbf{q}, \mathbf{p}) = 2\pi m$, $m \in \mathbb{Z}$.  The lattice summation energy is the total energy per particle for a given lattice, defined mathematically by:
\begin{equation}
\Phi(\mathbf{x}_1,\dotsc,\mathbf{x}_N) = \left(\frac{N}{2}\right)\sum_{j=2}^N \exp\left[-\left(\xi_j a\right)^2\right]\label{solid5},
\end{equation}
where the $a$ is the nearest-neighbor distance within the lattice and $\xi_j$ is a scaling factor that identifies the position of particle $j$ relative to particle 1.  Passing to the thermodynamic limit and partitioning the summation in \eqref{solid5} such that it is over all coordination shells in the lattice gives the desired result:
\begin{equation}\label{solid6}
\Phi/N = \frac{1}{2}\sum_{\nu = 1}^{+\infty} Z_{\nu} \exp\left[-(\xi_{\nu} a)^2\right].
\end{equation}
The factor $Z_{\nu}$ denotes the coordination number of the $\nu$-th coordination shell in the lattice.  We note that the nearest-neighbor distance to the power $d$ is inversely related to the density (i.e., $a^d \rho = c_{\Lambda}$), and the constant of proportionality depends on the chosen lattice.  Therefore, all that is needed to completely specify the lattice summation energy for a given lattice are the sets $\{Z_{\nu}\}$ and $\{\xi_{\nu}\}$ along with the proportionality constant $c_{\Lambda}$.  We consider here the integer lattices $\mathbb{Z}^d$ and the lattice families $D_d, A_d,$ and $E_d$ along with their respective duals.  The $D_d$ lattices are the $d$-dimensional counterparts to the three-dimensional FCC lattice and are the densest known lattice packings for all $d < 6$; similarly, the $A_d$ lattice family generalizes the $d = 2$ triangular lattice.  The $E_d$ family contains the densest known lattice packings for $d = 6, 7, 8$ but are not uniquely defined for $d < 6$.  

\subsection{The fluid-solid phase transition in the high-density limit}

From our knowledge of the three-dimensional phase diagram of the GCM, we expect to find a fluid-solid phase transition in any dimension $d$ such that $T_m \rightarrow 0$ in the limits $\rho \rightarrow 0$ and $\rho \rightarrow +\infty$.  The former limit is expected by the reduction of the GCM to a system of hard spheres in this regime, which has been shown with generality by Stillinger.\cite{St76}  There is strong numerical support (though no rigorous proof) for a fluid-solid phase transition with hard spheres up to $d = 3$, and we strongly suspect this is still true for higher dimensions.  For $d = 3$, the freezing temperature for a hard-sphere system scales as  \cite{LaLiWaLoe00} $T_f (\rho) \sim \exp\left(-c \rho^{-2/3}\right)$, where $c$ is a constant, and we conjecture that for arbitrary $d$ the scaling is similar.  In any case, our focus here is on the limit $\rho \rightarrow +\infty$.  We initially consider a finite system of $N$ particles in a volume $\Omega$ and introduce so-called \emph{collective coordinates}, defined such that:
\begin{align}
\Phi(\mathbf{x}_1, \dotsc, \mathbf{x}_N) &= \sum_{1 \leq i < j \leq N} \phi(r_{ij})\label{rsolid1}\\
&= \left(\frac{1}{2\Omega}\right)\sum_{\mathbf{k}} \hat{\phi}_{\Omega}(\mathbf{k})\left[\rho(\mathbf{k})\rho(-\mathbf{k}) - N\right]\label{rsolid2},
\end{align}
where:
\begin{align}
\hat{\phi}_{\Omega}(\mathbf{k}) &= \int_{\Omega} \phi(\lVert \mathbf{x}\rVert) \exp\left[- i (\mathbf{k}, \mathbf{x})\right]d\mathbf{x}\label{rsolid3}\\
\rho(\mathbf{k}) &= \sum_{i=1}^N \exp\left[-i (\mathbf{k}, \mathbf{x}_i)\right]\label{rsolid4}.
\end{align}
In passing to the thermodynamic limit, we have that $\hat{\phi}_{\Omega} \rightarrow \hat{\phi}$, which is given in \eqref{intro3}.  

Let $k_0$ be defined as a reciprocal space radius such that $\beta\hat{\phi}(k_0) = \frac{1}{2}$; therefore,
\begin{equation}\label{rsolid5}
k_0 = 2\sqrt{\ln\left(2\pi^{d/2}\beta\right)}.
\end{equation}
The results in \eqref{intro3} imply that $\hat{\phi}(k)$ decreases as $k$ increases.  As a result, in the low-temperature regime for $\lVert\mathbf{k}\rVert < k_0$, \eqref{rsolid5} implies that $\beta\hat{\phi}(k)$ is large and positive.  By \eqref{rsolid2}, minimization of the energy will then require $\rho(\mathbf{k})\rho(\mathbf{-k})$ to move toward its minimum to offset the effect of increasing $\hat{\phi}(k)$.  However, for $\lVert\mathbf{k}\rVert > k_0$ $\beta\hat{\phi}(k)$ will become increasingly small, and the magnitude of $\rho(\mathbf{k})\rho(\mathbf{-k})$ is of less consequence in the minimization of $\Phi$.  

Since the density of $k$-vectors inside a sphere of volume $\Omega$ is $\Omega/(2\pi)^d$, the number $N_0$ of such vectors with magnitude $k \leq k_0$ is:
\begin{equation}\label{rsolid6}
N_0 = \left(\frac{\pi^{d/2} k_0^d}{\Gamma(d/2 + 1)}\right)\left(\frac{\Omega}{(2\pi)^d}\right) = \frac{k_0^d \Omega}{2^d \Gamma(d/2 + 1) \pi^{d/2}}.
\end{equation}
By the argument above, we conclude that for sufficiently high density $\rho$, $N_0$ reflects the number of ``lost'' degrees of freedom to the system by the argument above.  When $N_0$ reaches some characteristic fraction $0 < \theta < 1$ of the total number $Nd$ of degrees of freedom, the GCM will freeze; i.e.:
\begin{equation}\label{rsolid7}
\theta Nd = \frac{\left(\ln\left[2\beta_f \pi^{d/2}\right]\right)^{d/2} \Omega}{\pi^{d/2}\Gamma[d/2 + 1]},
\end{equation}
where $\beta_f$ is the reciprocal freezing temperature.  Solving \eqref{rsolid7} for $k_B T_f$ yields:
\begin{equation}\label{rsolid8}
k_B T_f = 2 \pi^{d/2} \exp\left[-\left(d\theta \rho \pi^{d/2}\Gamma[d/2 + 1]\right)^{2/d}\right],
\end{equation}
which approaches 0 in the limit $\rho \rightarrow +\infty$. 

\subsection{Duality relationships}

Since the Gaussian is self-similar under Fourier transform, it is possible to relate the lattice summation energy of a lattice at low density to the lattice summation energy of its dual lattice at high density; we call such an expression a type of \emph{duality relation}.  A dimensionally-dependent duality relation derived in this study for the lattice summation energies in the GCM is given below; however, we also mention the duality relationships recently put forth by Torquato and Stillinger regarding the ground state of a classical system interacting via a bounded, absolutely integrable pair potential $\phi(r)$.\cite{ToSt07}  One result is (based on the FT convention used in \eqref{intro2a}):
\begin{equation}\label{solid7}
\int_{\mathbb{R}^d}\phi(r)h(r) d\mathbf{r} = \left(\frac{1}{2\pi}\right)^{d}\int_{\mathbb{R}^d}\hat{\phi}(k)\hat{h}(k) d\mathbf{k},
\end{equation}
where $h$ denotes the total correlation function as defined in \eqref{dilute8}, and $\hat{h}, \hat{\phi}$ denote the Fourier transforms of $h$ and $\phi$, respectively.  Equation \eqref{solid7} is an immediate consequence of Parseval's formula (for a reference, see Lieb/Loss \cite{LiLo}) since under the given assumptions $\phi, h \in L^2(\mathbb{R}^d)$; furthermore, one may show that if the configuration of particles in $\mathbb{R}^d$ is a ground state and ergodicity is assumed, then the left- and right-hand sides of \eqref{solid7} are minimized.  It should be stressed, however, that \eqref{solid7} will hold regardless of whether the configuration is a ground state.  Equation \eqref{solid7} may in turn be used to prove the following duality relationship for a Bravais lattice $\Lambda$:\cite{ToSt07}
\begin{equation}\label{solid8}
\phi(r = 0) + \sum_{\mathbf{r} \in \Lambda\setminus\{\mathbf{0}\}} \phi(r) = \rho \hat{\phi}(k = 0) + \rho \sum_{\mathbf{k} \in \Lambda^*\setminus\{\mathbf{0}\}} \hat{\phi}(k),
\end{equation}
which follows from the identity $h(r) = \frac{1}{\rho s_1(r)} \sum\limits_{n=1} Z_n \delta(r-r_n) - 1$ for a Bravais lattice, where $s_1 (r)$ denotes the surface area of a $d$-dimensional sphere.  

Torquato and Stillinger go on to show that twice the minimized energy per particle $\hat{\Phi}_{\text{min}}$ for any ground-state structure of the dual potential $\hat{\phi}(k)$ is bounded from above by the corresponding real-space minimized twice-energy per particle $\Phi_{\text{min}}$, i.e., the right-hand side of \eqref{solid8}:
\begin{equation}\label{solid9}
\hat{\Phi}_{\text{min}} \leq \Phi_{\text{min}} = \rho \hat{\phi}(k = 0) + \rho \sum_{\mathbf{k} \in \Lambda^*\setminus \{\mathbf{0}\}} \hat{\phi}(k).
\end{equation}
This inequality results from the notion that the energy-minimizing configuration in the dual space to a real-space configuration need not be a Bravais lattice.  At the very least, such a possibility cannot be eliminated solely from \eqref{solid8}.  However, equality of minimum energies in real and reciprocal spaces will hold whenever the reciprocal lattice $\Lambda^*$ at reciprocal lattice density $\hat{\rho} = \rho^{-1} (2\pi)^{-d}$ is a ground state of $\hat{\phi}(k)$.  Conversely, if a sufficiently well-behaved dual potential $\hat{\phi}(k)$ has a Bravais lattice $\Lambda^*$ at number density $\hat{\rho}$, then:
\begin{equation}\label{solid10}
\Phi_{\text{min}} \leq \hat{\Phi}_{\text{min}} = \hat{\rho} \phi(r = 0) + \hat{\rho}\sum_{\mathbf{r} \in \Lambda\setminus\{\mathbf{0}\}} \phi(r).
\end{equation}

Here we present a duality relationship that associates the energy per particle $(\Phi/N)_\Lambda$ of a given lattice $\Lambda$ in the thermodynamic limit at low density with the equivalent energy per particle of the dual lattice $\Lambda^*$ at high density.  In accordance with prior work by Stillinger \cite{St79b} concerning one, two, and three dimensional duality relations for the GCM, we consider the energy per particle $(\Phi/N)$ to eliminate boundary effects in passing to the thermodynamic limit.  
Define:
\begin{equation}\label{rsolid9}
I_{\Lambda}(a) = 1+ \lim_{N\rightarrow +\infty} \left(\frac{2\Phi}{N}\right)_{\Lambda},
\end{equation}
where $a$ denotes the nearest-neighbor distance within the lattice.  We may equivalently write $I_{\Lambda}(a)$ in terms of the discrete density function $\varrho(\mathbf{s})$:
\begin{equation}\label{rsolid14}
I_{\Lambda}(a) = \int_{\mathbb{R}^d} \varrho(\mathbf{s}) \exp(-s^2) d\mathbf{s},
\end{equation}
where $\varrho(\mathbf{s})$ is given in terms of a summation over Dirac delta functions:
\begin{equation}\label{rsolid15}
\varrho(\mathbf{s}) = \sum_j \delta^{(d)} (\mathbf{s}-\mathbf{s}_j).
\end{equation}
It is convenient at this point to ``smooth'' the Dirac delta functions in \eqref{rsolid15} via convolution with a normalized Gaussian, yielding:
\begin{align}
\varrho(\mathbf{s}) &= \lim_{\alpha \rightarrow +\infty} \varrho(\mathbf{s}, \alpha)\label{rsolid16}\\
\varrho(\mathbf{s}, \alpha) &= \left(\frac{\alpha}{\pi}\right)^{d/2} \sum_j \exp\left[-\alpha(\mathbf{s}-\mathbf{s}_j)^2\right]\label{rsolid17},
\end{align}
where it is understood that the limit $\alpha \rightarrow +\infty$ is to be taken at an appropriate point in the calculation.  
Noting that $\varrho(\mathbf{s}, \alpha)$ is a $d^{th}$-order periodic function of the variable $\mathbf{s}$, we may represent this function in a Fourier series:
\begin{equation}\label{rsolid18}
\varrho(\mathbf{s}, \alpha) = \sum_{\mathbf{k}} f(\mathbf{k}) \exp\left[i (\mathbf{k}, \mathbf{s})\right]
\end{equation}
where $(\mathbf{x}, \mathbf{y}) = \sum_{i=1}^{d} x_i y_i$ denotes the inner product of two vectors in $d$-dimensional real Euclidean space, and the vectors $\mathbf{k}$ are $2\pi$ times the vectors from the dual lattice $\Lambda^*$.
Expressions for each $f(\mathbf{k})$ are determined from Fourier orthogonality conditions, whereby one multiplies \eqref{rsolid18} by $\exp\left[-i(\mathbf{k}, \mathbf{s})\right]$ and integrates over a unit cell within the lattice to obtain:
\begin{equation}\label{rsolid19}
f(\mathbf{k}) = \rho_{\Lambda}(a) \exp\left[\frac{-k^2}{4\alpha}\right],
\end{equation}
where $\rho_{\Lambda}(a) = c_{\Lambda}/a^d$ denotes the density of the system as a function of the nearest-neighbor distance $a$ with $c_{\Lambda}$ a constant for the lattice $\Lambda$.
Combining \eqref{rsolid16}-\eqref{rsolid19} yields:
\begin{align}
I_{\Lambda}(a) &= \lim_{\alpha\rightarrow +\infty} \int_{\mathbb{R}^d} \sum_{\mathbf{k}} f(\mathbf{k}) \exp\left[i(\mathbf{k},\mathbf{s})\right] \exp(-s^2) d\mathbf{s}\label{rsolid20}\\
&= \lim_{\alpha\rightarrow +\infty} \sum_{\mathbf{k}} f(\mathbf{k}) \int_{\mathbb{R}^d} \exp\left[i (\mathbf{k}, \mathbf{s})\right] \exp(-s^2) d\mathbf{s}\label{rsolid21}\\
&= \pi^{d/2}\rho_{\Lambda}(a) \lim_{\alpha\rightarrow +\infty}\sum_{\mathbf{k}} \exp\left[\frac{-k^2(\alpha+1)}{4\alpha}\right]\label{rsolid22}\\
&= \pi^{d/2}\rho_{\Lambda}(a)\sum_{\mathbf{k}}\exp\left[\frac{-k^2}{4}\right]\label{rsolid23}.
\end{align}
The right-hand side of \eqref{rsolid23} is exactly of the form for $I(a)$ given in \eqref{rsolid9}; in fact, under suitable scaling for the dual lattice $\Lambda^*$, we may write:
\begin{equation}\label{rsolid24}
I_{\Lambda}(a) = \pi^{d/2} \rho_{\Lambda}(a) \sum_{\mathbf{k}}\exp\left\{-\left[\pi\chi_{\mathbf{k}}b(a)\right]^2\right\},
\end{equation}
where $b(a)$ denotes the nearest-neighbor distance for the dual lattice as a function of $a$, and $\chi_{\mathbf{k}}$ denotes the related scaling factor for the particle coordinate in the dual lattice.  Reference to \eqref{rsolid9} yields the desired result:
\begin{equation}\label{rsolid25}
I_{\Lambda}(a) = \pi^{d/2}\rho_{\Lambda}(a)~I_{\Lambda^*}[\pi b(a)].
\end{equation}

Our interest here is to discern the so-called self-dual density $\bar{\rho}^*$, which is the density at which the lattice $\Lambda$ and its dual $\Lambda^*$ have the same energy per particle in the thermodynamic limit.  It is immediately clear from \eqref{rsolid25} that if the condition of equal energy per particle between a lattice and its dual holds, then:
\begin{equation}\label{rsolid26}
\bar{\rho}^* = \pi^{-d/2},
\end{equation}
whereby the coefficient on the right-hand side of \eqref{rsolid25} becomes unity.  

\subsection{Lattice energies and coexistence regions}

\begin{figure*}
\includegraphics[width=\columnwidth]{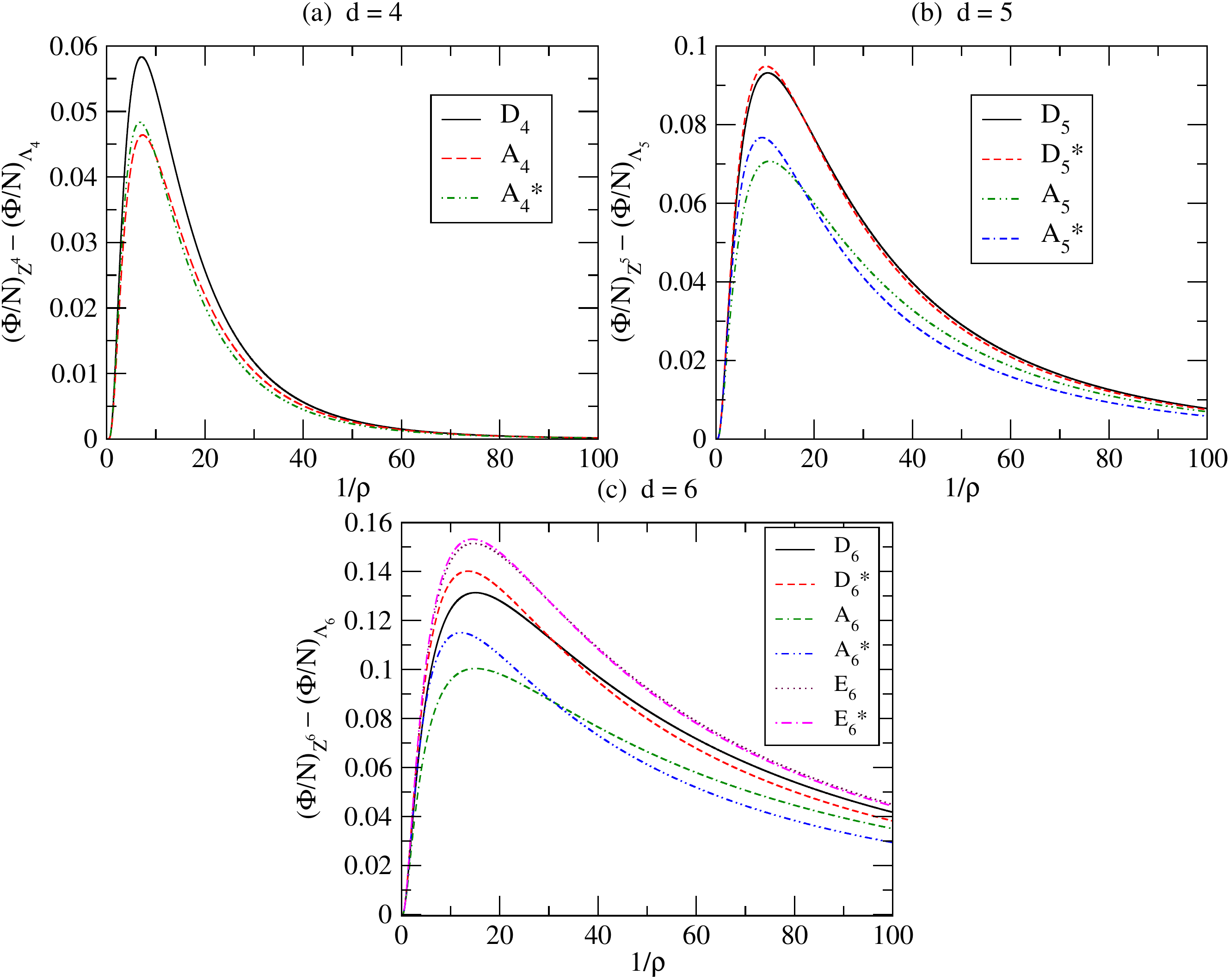}
\caption{\label{figfifteen}Relative lattice summation energies $\left(\Phi/N\right)_{\mathbb{Z}^d} - \left(\Phi/N\right)_{\Lambda_d}$ v. $1/\rho$ for given lattice families $\Lambda_d$; $d = 4, 5, 6$.}
\end{figure*}

\begin{figure*}
\includegraphics[width=\columnwidth]{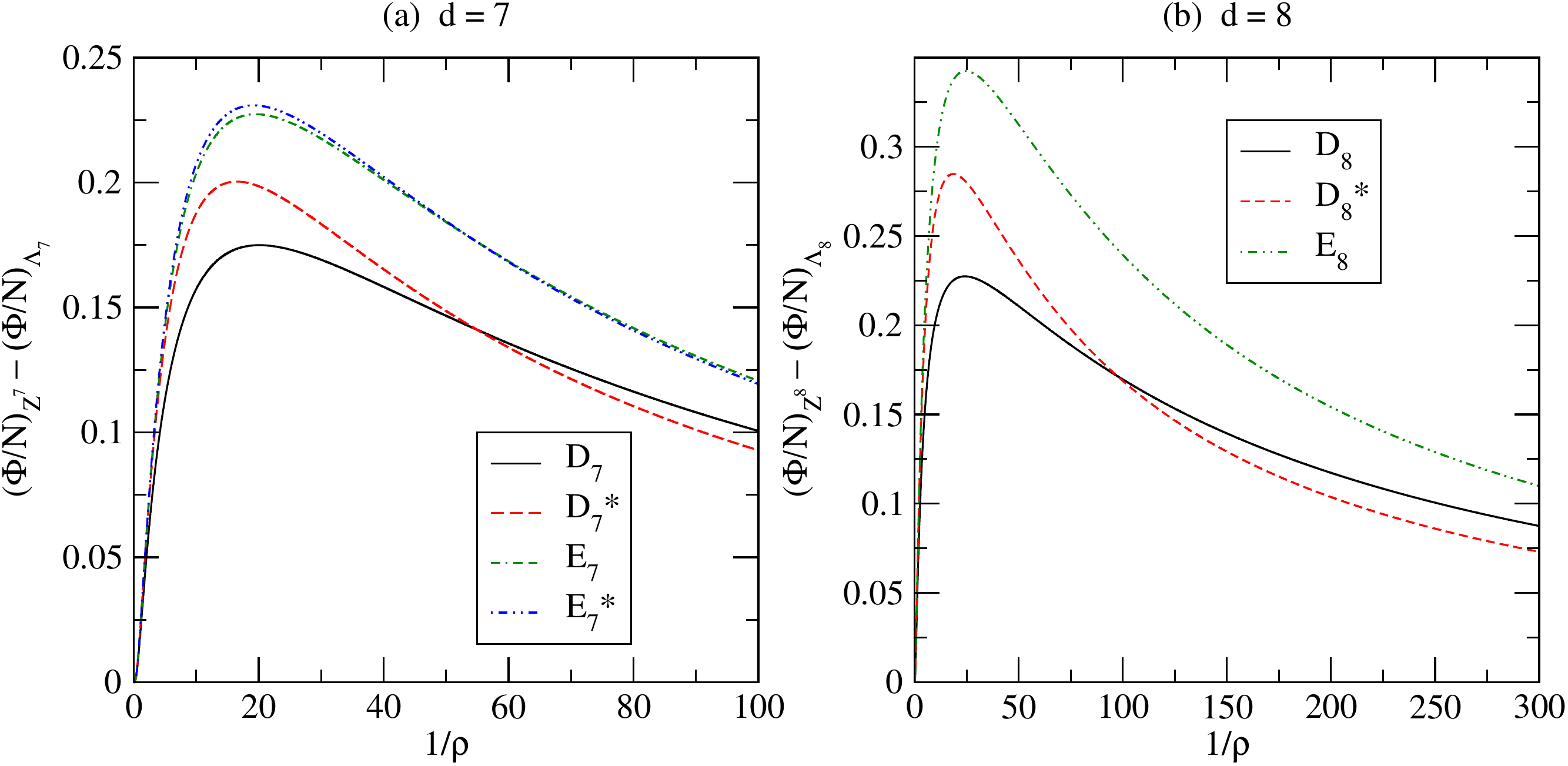}
\caption{\label{figsixteen}Relative lattice summation energies $\left(\Phi/N\right)_{\mathbb{Z}^d} - \left(\Phi/N\right)_{\Lambda_d}$ v. $1/\rho$ for given lattice families $\Lambda_d$; $d = 7, 8$.}
\end{figure*}

The lattice summation energies for our chosen lattice families ($D_d, A_d, E_d$) relative to the energy for the corresponding $\mathbb{Z}^d$ lattice are given in Figures \ref{figfifteen} and \ref{figsixteen}.  In accordance with our predictions from the duality relationship in \eqref{rsolid25}, there appears to be a universal phase-transition density between and lattice and its dual at $\rho = \pi^{-d/2}$.  For $d = 4, 5$, the lowest-energy lattices are given by the $D_d$ lattice and its dual; this observation provides direct support for the Torquato-Stillinger conjecture\cite{ToSt07} concerning the ground-state structures of classical systems since $D_d$ is the Bravais lattice corresponding to the densest known sphere packing for $3 \leq d \leq 5$.  As we would predict from this conjecture, the $E_d$ lattices and their duals obtain the lowest energy for $6 \leq d \leq 8$, followed by the $D_d$ family.  Thus, there appears to be a relationship between the density of a given lattice structure and its energy with respect to the GCM, providing numerical support for the conjecture mentioned above.    


We may define the lattice coexistence region between dual lattices via the Maxwell double-tangent construction.  For notational convenience, we will define:
\begin{align}
\varepsilon &= (\Phi/N)\label{solid11}\\
\bar{v} &= (1/\rho) = (V/N)\label{solid12}.
\end{align}
For a one-component system, the internal energy $\Phi$ satisfies the following differential:
\begin{equation}\label{solid13}
d\Phi = TdS-pdV+\mu dN,
\end{equation}
where $\mu$ is the chemical potential of the system.  Since we will be interested here in the variation of the internal energy with reciprocal density, we will without loss of generality assume constant entropy and simply write:
\begin{equation}\label{solid14}
d\Phi = -pdV+\mu dN.
\end{equation}

The Maxwell double-tangent construction involves finding a solution $(\bar{v}^{(\Lambda)}, \bar{v}^{(\Lambda^*)})$ to the following set of coupled equations:
\begin{align}
p^{(\Lambda)}(\bar{v}^{(\Lambda)}) &= p^{(\Lambda^*)}(\bar{v}^{(\Lambda^*)})\label{solid15}\\
\mu^{(\Lambda)}(\bar{v}^{(\Lambda)}) &= \mu^{(\Lambda^*)}(\bar{v}^{(\Lambda^*)})\label{solid16},
\end{align}
where $\bar{v}^{(\Lambda)}$ and $\bar{v}^{(\Lambda^*)}$ characterize the upper and lower reciprocal density bounds for the phase coexistence region between a lattice and its dual.  That \eqref{solid15} and \eqref{solid16} correspond to a double-tangent is an immediate consequence of Euler's theorem for first-order homogeneous functions (a succinct review may be found in Chandler \cite{Ch}) and \eqref{solid14}, which imply:
\begin{equation}\label{solid17}
\Phi = -pV + \mu N.
\end{equation}
Therefore, \eqref{solid16} is equivalent to:
\begin{equation}\label{solid18}
\left(\varepsilon+p\bar{v}\right)^{(\Lambda)} = \left(\varepsilon + p\bar{v}\right)^{(\Lambda^*)}.
\end{equation}
The equal-pressure condition in \eqref{solid15} allows us to write the following equation for a line tangent to $\varepsilon^{(\Lambda)}$ and $\varepsilon^{(\Lambda^*)}$ at $\bar{v}^{(\Lambda)}$ and $\bar{v}^{(\Lambda^*)}$, respectively:
\begin{equation}\label{solid19}
\varepsilon^{(\Lambda)}-\varepsilon^{(\Lambda^*)} = -p\left[\bar{v}^{(\Lambda)}-\bar{v}^{(\Lambda^*)}\right].
\end{equation}

To solve for $\bar{v}^{(\Lambda)}$ and $\bar{v}^{(\Lambda^*)}$, we utilize \eqref{solid14} to obtain:
\begin{align}
p &= -\left(\frac{\partial\Phi}{\partial V}\right)_N = -\left(\frac{\partial \varepsilon}{\partial \bar{v}}\right)\label{solid20}\\
\mu &= \left(\frac{\partial\Phi}{\partial N}\right)_V = \left(\frac{\partial (N\varepsilon)}{\partial N}\right)_V = \varepsilon + N \left(\frac{\partial\varepsilon}{\partial N}\right)_V\\ 
&= \varepsilon + N\left(\frac{\partial\varepsilon}{\partial\bar{v}}\right)\left(\frac{\partial\bar{v}}{\partial N}\right)_V = \varepsilon-\bar{v}\left(\frac{\partial\varepsilon}{\partial\bar{v}}\right)\\ 
&= \varepsilon+\bar{v}p\label{solid21}.
\end{align}
Equations \eqref{solid20} and \eqref{solid21} allow us to numerically determine values for $\bar{v}^{(\Lambda)}$ and $\bar{v}^{(\Lambda^*)}$ from our lattice summation data.

Table \ref{coextab1} collects the results for the Maxwell construction of the phase-coexistence regions between dual lattices in each dimension.  We note that the width of the phase coexistence region scaled by the self-dual density $\bar{\rho}^* = \pi^{-d/2}$ increases with dimension for a given lattice family.  The immediate significance of this behavior is that for arbitrarily high Euclidean dimensions, it is possible that the phase coexistence region for a particular lattice is wide enough such that new structures are able to achieve lower energy as ground states.  In the context of a decorrelation principle, we cannot exclude the possibility that these structures are disordered.  If the conjecture by Torquato and Stillinger \cite{ToSt06} that the densest known packings of hard spheres in high dimensions are disordered is to be believed, then this behavior is expected.  

\begin{table*}[!htp]
\begin{ruledtabular}
\begin{tabular}{c c c c c c c}
$d$& 
Lattice Family& 
$\mu^*$& 
$p^*$& 
$\rho_{\Lambda}$ & $\rho_{\Lambda^*}$&
$\lvert \rho_{\Lambda}-\rho_{\Lambda^*}\rvert/\bar{\rho}^*$\\
\hline
4 & $A$ & 0.47983 & 0.03310 & 0.10103 & 0.10164 & 0.00605\\
5 & $A$ & 0.53922 & 0.01980 & 0.05683 & 0.05749 & 0.01154\\
5 & $D$ & 0.51155 & 0.01927 & 0.05707 & 0.05726 & 0.00332\\
6 & $A$ & 0.60331 & 0.01186 & 0.03196 & 0.03254 & 0.01783\\
6 & $D$ & 0.56598 & 0.01146 & 0.03206 & 0.03244 & 0.01154\\
6 & $E$ & 0.54849 & 0.01136 & 0.03236 & 0.03244 & 0.00248\\
7 & $D$ & 0.62835 & 0.00684 & 0.01800 & 0.01841 & 0.02257\\
7 & $E$ & 0.57723 & 0.01826 & 0.01820 & 0.01826 & 0.00330\\
8 & $D$ & 0.69825 & 0.00410 & 0.01009 & 0.01045 & 0.03483
\end{tabular}
\end{ruledtabular}
\caption{\label{coextab1}Scaled phase coexistence regions for specified lattices in 4, 5, 6, and 7 dimensions.  Here $\mu^*$ and $p^*$ are the self-dual chemical potential and pressure, respectively; $\bar{\rho}^*$ is the self-dual density as defined in \eqref{rsolid26} above.}
\end{table*}

\section{Concluding Remarks}

We have hereby made an effort to generalize the phase properties of the GCM up to $d = 3$ with respect to the dimensionality of the system.  In the fluid phase, we have developed a dimensionally-dependent low-density/high-temperature expansion for the radial distribution function $g_2$.  Although this series provides some interesting evidence for a decorrelation principle, it suffers from the drawback of being divergent due to a collapse instability induced by the unmediated attraction of the particles for values of $\beta < 0$.  Numerical approximations (HNC, PY, and MFA) for $g_2$ at higher densities and lower temperatures strengthen the evidence for decorrelation in the GCM, and we show that the MFA contains the high-dimensional behavior of the GCM for sufficiently high temperatures.  With regard to the solid phase of the GCM, results for the lattice summation energies of known lattice families up to $d = 8$ provide direct support for the recent Torquato-Stillinger conjecture \cite{ToSt07} concerning the ground states of classical many-particle systems.  Namely, for sufficiently well-behaved pair potentials the ground states correspond to the densest known lattice packings for low densities and the corresponding dual lattices for high densities with a solid-solid phase transition at intermediate density values.  Having mentioned this idea, we cannot exclude the possibility that for sufficiently high dimensions the ground-state structures may in fact be disordered; in the case where the particles adhere to a decorrelation principle, it is not unreasonable that we might expect this result in accordance with prior work by Torquato and Stillinger.\cite{ToSt06}
 
Despite our current work on this problem, a few remaining points are worthy of mention.  With regard to the behavior of the system in the fluid regime, the nature of the collapse instability makes analytical evaluation of the radial distribution function difficult to interpret.  We have developed the density expansion of $g_2$ and made favorable comparisons to numerical approximations, yet the physicality of the cluster expansion seems to be lost due to divergence of the series.  Stillinger \cite{St79} priorly utilized a Borel resummation to interpret the ``lost'' information in the high-temperature expansion of the excess free energy, yet the strong spatial and dimensional dependence of the $g_2$ expansion makes this technique difficult in the present case.  What is perhaps most promising is the reduction of the expansion to that of the MFA in the high-dimensional limit, whereby the divergence of the series is partially removed with results similar to the numerical approximations from the HNC.  However, the MFA has been shown to be a thermodynamically inconsistent approximation,\cite{LiMlGoKa07} meaning that it cannot be an exact theory to describe the GCM.

In the solid phase, we still have an open question concerning the fluid-solid phase transition at any value of $\rho$.  There is reason to believe, as mentioned above, that such a transition exists up to some maximum melting temperature $T_m(\rho)$ as in the three-dimensional case,\cite{LaLiWaLoe00} and we have implicitly made this assumption in the present work.  However, we know that the phase diagram will be dimensionally-dependent as shown by the fact that for $d = 4$, the self-duality of $D_4$ preempts a solid-solid phase transition as in $d = 3$.  Nevertheless, the study of the lattice structures of the GCM is worthwhile in the context of the Torquato-Stillinger conjecture\cite{ToSt07} concerning the ground states of classical systems and with respect to the decorrelation principle.  Though the proof of the existence of disordered ground states for the GCM in high dimensions is still unavailable, our results show that we certainly cannot preclude the possibility.

The results we have presented suggest that it may be possible to extend this research to a broader class of pair interactions.  We recall that the primary advantage of the GCM, as we have shown, is the property of being self-similar under Fourier transform.  However, it is known that the eigenfunctions of the Fourier transform are equivalent to the states of the quantum-mechanical harmonic oscillator (under suitable scaling), namely Gaussian functions modulated by the Hermite polynomials.  This property indicates that our analysis may be generalized to pair potentials containing both repulsive and attractive components.  This more general case is likely to modify the collapse instability, and determining the precise manner of this modification is an area that deserves further study.  Physically, these interactions could be related to spatially inhomogeneous solvent compositions that simultaneously induce repulsion and attraction among macromolecules in solution.  Alternatively, one may also consider pair potentials formed from linear combinations of (attractive and repulsive) Gaussian interactions.  In any case, the possibility for finding unique thermodynamic phenomena which could be used in the design of novel materials makes this a viable avenue for future exploration. 

\begin{acknowledgments}
We benefited greatly from discussions with Henry Cohn. 
This work was supported by the Office of Basic Energy Sciences, US Department of Energy, under Grant DE-FG02-04-ER46108. 
\end{acknowledgments}

\appendix

\section{Proof of regularity of the GCM pair potential}

\textbf{Proposition:}  \textit{The GCM pair potential is regular for $\epsilon > 0$.}

\textit{Proof:}   From the definition of regularity, it is sufficient to prove the proposition for $\beta > 0$.  Fix $\beta, \epsilon > 0$.  Positivity of the pair potential \eqref{intro2} provides the requisite lower bound of $K = B = 0$, where $B$ denotes the stability constant for the pair interaction.  Since $0 < \exp\left[-\beta \phi(\mathbf{x})\right] \leq 1$ for $\beta, \epsilon > 0$, we thus have:
\begin{equation}\label{appendixa1}
C(\beta) = \int_{\mathbb{R}^d}\left\{1-\exp\left[-\beta^* \exp\left(-\alpha \mathbf{x}^2\right)\right]\right\}d\mathbf{x},
\end{equation}
where $\alpha = 1/\sigma^2$ and $\beta^* = \beta \epsilon > 0$.

Since the exponential function is an entire function, we may write the integrand of \eqref{appendixa1} as a uniformly convergent Taylor series:
\begin{equation}\label{appendixa2}
C(\beta) = \int_{\mathbb{R}^d} \sum_{n=1}^{+\infty} \left\{\frac{(-1)^{n+1}(\beta^*)^n}{n!} \exp\left[-\alpha n \mathbf{x}^2\right]\right\} d\mathbf{x}.
\end{equation}

The series in \eqref{appendixa2} is uniformly convergent on any compact interval in $\mathbb{R}$, which implies we may exchange the operations of integration and summation to obtain:
\begin{align}
C(\beta) &= \sum_{n=1}^{+\infty}\left\{\frac{(-1)^{n+1}(\beta^*)^n}{n!} \int_{\mathbb{R}^d} \exp\left[-\alpha n \mathbf{x}^2\right] d\mathbf{x}\right\}\label{appendixa3}\\
&= \sum_{n=1}^{+\infty} \frac{(-1)^{n+1}(\beta^*)^n}{n!} \left(\frac{\pi}{\alpha n}\right)^{d/2}\label{appendixa4}.
\end{align}

One may show that the series in \eqref{appendixa4} is convergent via reference to the ratio test.  
\begin{align}
\lim_{n\rightarrow +\infty} \zeta_n &= \lim_{n\rightarrow +\infty} \left\lvert\left[\frac{(\beta^*)^{n+1}}{(\beta^*)^{n}}\right] \cdot \left[\frac{n!}{(n+1)!}\right] \cdot \left(\frac{n}{n+1}\right)^{d/2}\right\rvert\label{appendixa5}\\
&= \beta^* \lim_{n\rightarrow +\infty} \left\{\left(\frac{1}{n+1}\right)\left(\frac{n}{n+1}\right)^{d/2}\right\}\label{twentysixh}\\
&= 0 < 1\label{appendixa6}.
\end{align}

Convergence of the series in \eqref{appendixa4} proves the proposition. $\square$

\section{Proof of Equation (32)}

\textbf{Proposition:}  The chain diagrams defined by \eqref{rdilute6} satisfy:
\begin{equation}\label{appendixb1}
C_{n+1} = \rho^{n-1}\left(\frac{\pi^{n-1}}{n}\right)^{d/2} \phi^{1/n}
\end{equation}
$\forall n \geq 2$, where $\phi$ denotes the GCM pair potential.

\textit{Proof:}  We prove by mathematical induction.  Let $p(k), k \in \mathbb{N}\setminus \{1\}$, be the claim:
\begin{equation}\label{appendixb2}
\int \prod_{i=1}^{k} \phi_{i,(i+1)} d\mathbf{x}_2 \dotsm d\mathbf{x}_k = \left(\frac{\pi^{k-1}}{k}\right)^{d/2}\phi_{1,(k+1)}^{1/k}.
\end{equation}
As mentioned in the text, we have chosen the stationary vertices in the cluster diagram to be $1$ and $k+1$ for simplicity.  To prove $p(2)$, we use the relation:
\begin{equation}\label{appendixb3}
\int_{\mathbb{R}^d} \phi_{i,j}^m \phi_{j,k}^n d\mathbf{x}_j = \left(\frac{\pi}{m+n}\right)^{d/2} \phi_{i,k}^{\frac{mn}{m+n}},
\end{equation}
which follows directly from \eqref{intro4}.

Equation \eqref{appendixb3} implies:
\begin{equation}\label{appendixb4}
\int \phi_{1,2} \phi_{2,3} d\mathbf{x}_2 = \left(\frac{\pi}{2}\right)^{d/2} \phi_{1,3}^{1/2},
\end{equation}
which is the exact form of the right-hand side of \eqref{appendixb1}.  Therefore, $p(2)$ is true.

Fix $k_0 \in \mathbb{N}\setminus \{1\}$ arbitrarily, and make the usual induction hypothesis that $p(k_0)$ is true.  We now prove $p(k_0 + 1)$ is true.  Using Fubini's theorem:
\begin{align}
\int \prod_{i=1}^{k_0 + 1} \phi_{i,(i+1)} d\mathbf{x}_2 \dotsm d\mathbf{x}_{k_0 +1} &= \int \phi_{(k_0 + 1),(k_0 + 2)} \left(\prod_{i=1}^{k_0} \phi_{i,(i+1)} d\mathbf{x}_2 \dotsm d\mathbf{x}_{k_0}\right) \cdot d\mathbf{x}_{k_0 + 1}\label{appendixb5}\\
&= \left(\frac{\pi^{k_0-1}}{k_0}\right)^{d/2} \int \phi_{(k_0 + 1),(k_0 + 2)} \phi_{1,(k_0 + 1)}^{1/k_0} d\mathbf{x}_{k_0 + 1}\label{appendixb6}\\
&= \left(\frac{\pi^{k_0}}{k_0+1}\right)^{d/2} \phi_{1,(k_0 + 2)}^{1/(k_0 + 1)}\label{appendixb7}.
\end{align}

We see from the result in \eqref{appendixb7} that $p(k_0 + 1)$ is true, and the truth of $p(k)~\forall k \in \mathbb{N}\setminus \{1\}$ immediately follows.  The proof of \eqref{appendixb1} is thus apparent with the $n-1$ factors of $\rho$ arising by convention from the integration over the $n-1$ integrable nodes.  $\square$

\section{Picard iteration algorithm for numerical approximations}

In order to obtain a numerical approximation to $g_2$ using one of the closures mentioned in Section II, we are required to solve the OZ equation, which for a given closure $c(r)$ will be a nonlinear integral equation.  To address this problem, we first make an initial estimate for the convolution integral $\gamma(r) = h(r) - c(r)$ and calculate the corresponding $c(r)$ using either \eqref{dense3} or \eqref{dense4}.  To update our guess for the form of $\gamma(r)$, we use $\gamma^{\prime}(r) = \mathfrak{F}^{-1}\left\{\rho\left[\hat{c}(k)\right]^2/[1-\rho\hat{c}(k)]\right\}$, where $\mathfrak{F}^{-1}$ denotes the inverse Fourier transform, and the right-hand side of the equation follows from taking the FT of the OZ equation.  We calculate the error $\varepsilon = \sqrt{\delta r}\lVert\gamma^{\prime}(r) - \gamma(r)\rVert$, where $\delta r$ denotes the size of the coordinate-space mesh, and set $\gamma(r) = \alpha \gamma^{\prime}(r) + (1-\alpha)\gamma(r)$, where $0 < \alpha < 1$ is a mixing parameter to speed (or to aid) convergence of the algorithm.  The algorithm is iterated so long as $\varepsilon$ is greater than some specified tolerance.  Convergence of this algorithm is reasonably quick assuming a good choice of $\alpha$, requiring usually only a few hundred interations.  We note, however, that as the density increases, it is necessary to decrease the value of $\alpha$, thereby increasing the requisite iterations.  

\section{Proof of convergence of the series in (52)} 

\textbf{Proposition:}  The function $g_2^{(\text{MFA})}$ is well-defined by \eqref{rdense15}; more specifically,
\begin{equation}\label{appendixc1}
\sum_{n=2}^{+\infty} (-\beta)^n C_{n+1}(r) < +\infty
\end{equation}
$\forall r \in [0, +\infty)$ within some finite radius of convergence with respect to $\beta$.  

\textit{Proof:}  We show that the series in \eqref{appendixc1} converges by reference to the ratio test.  Namely,
\begin{align}
\zeta_{n+1}(r) &= \frac{C_{n+1}(r)}{C_n(r)}\label{appendixc2}\\
&= \left(\frac{\rho^n}{\rho^{n-1}}\right)\left(\frac{\pi^n}{\pi^{n-1}}\right)^{d/2} \left(\frac{n}{n+1}\right)^{d/2} \phi^{-\frac{1}{n(n+1)}}\label{appendixc3}\\
&\xrightarrow{n\rightarrow +\infty} \rho \pi^{d/2}\label{appendixc4}.
\end{align}
Therefore, the series in \eqref{appendixc1} will converge for all $\beta$ such that $\lvert\beta\rvert < 1/(\rho\pi^{d/2})$.  (Note that we have, without loss of generality, chosen unitless parameters as described in Section III.)  $\square$


\end{document}